\documentclass[pre,twocolumn,letterpaper,showpacs,floatfix]{revtex4}

\usepackage{graphicx,amsfonts,amsmath,amssymb,bm,latexsym,wasysym,setspace}

\begin{document}

\title{Model for Spreading of Liquid Monolayers}

\date{\today}

\author{M. N. Popescu}
\email{popescu@mf.mpg.de}
\author{S. Dietrich}
\email{dietrich@mf.mpg.de}

\affiliation{
Max-Planck-Institut f\"ur Metallforschung, Heisenbergstr. 3,
D-70569 Stuttgart, Germany,}
\affiliation{Institut f\"ur Theoretische und Angewandte Physik,
Universit\"at Stuttgart, Pfaffenwaldring 57, D-70569 Stuttgart,
Germany}

\begin{abstract}
Manipulating fluids at the nanoscale within networks of channels or
chemical lanes is a crucial challenge in developing small scale
devices to be used in microreactors or chemical sensors. In
this context, ultra-thin (i.e., monolayer) films, experimentally
observed in spreading of nano-droplets or upon extraction
from reservoirs in capillary rise geometries, represent an
extreme limit which is of physical and technological relevance
since the dynamics is governed solely by capillary forces.
In this work we use kinetic Monte Carlo (KMC) simulations to
analyze in detail a simple, but realistic model proposed by
Burlatsky \textit{et al.} \cite{Burlatsky_prl96,Oshanin_jml} for
the two-dimensional spreading on homogeneous substrates of a
fluid monolayer which is extracted from a reservoir.
Our simulations confirm the previously predicted time-dependence of
the spreading, $X(t \to \infty) = A \sqrt t$, with $X(t)$ as the
average position of the advancing edge at time $t$, and they reveal a
non-trivial dependence of the prefactor $A$ on the strength $U_0$
of inter-particle attraction and on the fluid density $C_0$ at the
reservoir as well as an $U_0$-dependent spatial structure of the
density profile of the monolayer. The asymptotic density profile at
long time and large spatial scale is carefully analyzed within the
continuum limit. We show that including the effect of correlations
in an effective manner into the standard mean-field description leads
to predictions both for the value of the threshold interaction above
which phase segregation occurs and for the density profiles in excellent
agreement with KMC simulations results.

\end{abstract}

\pacs{68.15.+e, 68.37.-d, 81.15.Aa}

\maketitle

\section{Introduction}

In the context of microfluidics, wetting phenomena at the micro- and
nano-meter scale \cite{deGennes_85,Leger_92,Oron_97,Dietrich_99,Voue_00}
are relevant for applications like microreactors or chemical sensors,
for which a crucial challenge is the transport of liquid to networks
of channels or chemical lanes, as well as its precise manipulation
within such a network \cite{Mitchell_01,Stone_01,Turner_02}.
Since at this small scale the liquid-substrate interaction is important,
the flow of thin films may be eventually controlled by engineering
the physico-chemical properties of the substrate, thus opening the
road for applications which do not have an equivalent at the macroscopic
scale \cite{Lindner_01,Zhao_01,Zhao_02}.

Although the existence of very thin precursor films has been long ago
evidenced by the studies of Hardy \cite{Hardy}, only
recent experiments on liquid spreading on atomically smooth surfaces
\cite{Heslot_jpcm,Heslot_89,Heslot_prl,Daillant_90,Leiderer_92,
Fraysse_jcis,Voue_lang,Perez},
performed with volumes of the order of nano-liters,
have clearly shown by means of dynamic ellipsometry or X-ray
reflectivity measurements that one or few precursor films with
\textit{molecular thickness} and \textit{macroscopic extent}
advance in front of the macroscopic liquid wedge of the spreading drop.
The liquids used were low-molecular-mass polymer oils which behave as
non-volatile liquids, and experiments performed both for spreading
of nano-droplets and for capillary rise geometries have established
that the linear extent $X(t)$ of the precursor film grows in time as
$X(t \to \infty) \sim A t^\alpha$. The exponent $\alpha = 1/2$ seems
to be independent of the nature of the liquid and of the substrate,
of the geometry, and of the volume of droplet as long as the droplet
is not emptied and constitutes a reservoir for the extracting film;
only the prefactor $A$ depends on these parameters.

Several theoretical models have been proposed (see Refs.
\cite{deGennes,Abraham_prl90,Kaski_95,Burlatsky_prl96,
Oshanin_jml,Abraham_02}
and references therein) and an impressive number of
Molecular Dynamics (MD) and Monte Carlo (MC) numerical simulations
have been performed (see Refs.
\cite{Kaski_92,Nissila_94,Koplik_96,Nissila_96,Herminghaus_96,
DeConinck_96,Bekink_96,Voue_00,Heine_03} and references therein)
in order to understand the mechanisms behind the extraction
of precursor films and to explain the time dependence of the
spreading.

The hydrodynamic model of de Gennes and Cazabat \cite{deGennes}
assumes a layered structure of the droplet, each layer being a
two-dimensional incompressible fluid, in which vertical transport
is possible only at the edges of the layers. The model leads to
the correct time dependence for the advancing layers, but, as
pointed out in Refs.~\cite{Burlatsky_prl96}, \cite{Abraham_prl90},
and \cite{Heine_03}, it is debatable if this hydrodynamic description
holds at the molecular level and can be directly applied to ultra-thin films.

A different approach, along the line of earlier work on activated
kinetics by Cherry, Holmes, Blake, and Haynes \cite{Cherry,Blake},
consists of a microscopic description for the thin liquid films in
terms of lattice gas models for interacting particles. One such
model is the two-dimensional driven Ising model recently proposed
by Abraham \textit{et al.} \cite{Abraham_02}. Using kinetic Monte
Carlo (KMC) simulations it was shown that in this model the transport
of mass occurs via a second layer, and a particle-hole diffusion
equation was used to show that the model leads to correct predictions
(confirmed also by simulations) for the time dependence of spreading.
The model predicts a uniform density and a compact first monolayer,
in close resemblance of the incompressible layers of the hydrodynamical
model mentioned before.

For the case that the precursor consists of a single monolayer, a
lattice gas model of interacting particles has been proposed by
Burlatsky \textit{et al.} \cite{Burlatsky_prl96}. This model, which
allows mass transport from the reservoir to the advancing edge only
inside the monolayer, has been  extended to the more general situation
of relaxation of a monolayer initially occupying a half-plane without
a reservoir by Oshanin \textit{et al.} \cite{Oshanin_jml}.
Based on several mean-field assumptions, among which the strongest
is the replacement of the inter-particles attraction in the ``bulk''
by an ``effective force'' acting on the advancing edge, the authors have
been able to derive the $t^{1/2}$-law for the spreading and to calculate
the dependence of the prefactor on the fluid-fluid interaction parameters.
In contrast to the other two models, in this case the density in the
monolayer depends significantly on the distance from the reservoir.
Although it is reasonable to expect that neglecting the attractive
particle-particle interactions in the ``bulk'' should not affect the
time dependence $\sim t^{1/2}$, one can expect that the behavior in
the presence of attractive interactions is much richer (see, for
example, the recent KMC simulations results of
Lacasta \textit{et al.} \cite{Lacasta} for a closely related
one-dimensional model).

All three models mentioned above recover the correct time-dependence
of spreading, but it is unlikely that one can discriminate between
them via experimental tests based on their predictions for the
corresponding prefactors because these include in each case a number
of parameters whose connections with experimental quantities are not
clear. However, we have already pointed out that these models lead to
qualitatively different predictions with respect to the shape of the
emerging density profiles (constant in the first two cases, spatially
varying in the last) which are, in principle, experimentally
accessible.

In this work we analyze in detail the density profiles of this last
model. We present results of KMC simulations on a square lattice of
a model for spreading of a liquid monolayer closely related to the
model in Refs.~\cite{Burlatsky_prl96} and \cite{Oshanin_jml}. Our
choice for KMC simulations of a lattice gas model is motivated by the
fact that we are interested in the asymptotic (large spatial and
temporal scales) behavior, a regime which as yet cannot be explored
using Molecular Dynamics simulations because of extensive computing
resources needed to simulate spatially large systems and unreasonable
large CPU times required to simulate real times even in the
order of $\mu$-seconds. In contrast to the previous work mentioned
above, we shall explicitely consider the asymmetry of the jump rates
in the bulk, at the expense of being able to measure the prefactor $A$
from the simulations but not to predict it analytically. Our results
show a non-trivial dependence of the prefactor $A$ on the strength
$U_0$ of the inter-particle attraction and on the density $C_0$ at
the reservoir. The asymptotic spreading behavior at long time and
large spatial scale of the transversally averaged density profile is
analyzed within a continuum limit. We show that the model predicts
qualitatively different structures for the experimentally accessible
density profiles along the spreading direction above and below a
threshold value for the ratio between the fluid-fluid interaction and
the thermal energy. Including the effect of correlations in an
effective manner into the standard mean-field description, we find
excellent agreement between the theoretical predictions and the KMC
results. We conclude the paper with a summary and discussion of the
results.

\section{Definition and Discussion of the Model}

As mentioned in the Introduction, a simple microscopic model for
the dynamics of a fluid monolayer in contact with a reservoir was
proposed in Refs. \cite{Burlatsky_prl96} and \cite{Oshanin_jml}.
Although we use here this lattice gas model of interacting particles
with only slight modifications, for clarity and further reference we
describe, motivate, and comment on the defining rules as follows.\\
%%%%%%%%%%%%%%%%
\textbf{(a)} The spreading geometry is rectangular ($x-y$ plane)
and the substrate is homogeneous. The half-plane $x < 0$ is occupied
by a reservoir of particles (three-dimensional bulk liquid) at fixed
chemical potential which maintains at its contact line with the
substrate, positioned at the line $x =0$, an {\it average} density
$C_0$ (defined as number of particles per unit length in the
transversal $y$ direction). For the case
of capillary rise, the reservoir would correspond to the liquid bath
and the line $x=0$ to the edge of the macroscopic meniscus. It is
assumed that the only role of the reservoir is to maintain $C_0$
constant, and thus to feed the monolayer which is extracted, but there
is no flow of particles from the reservoir to ``push'' the film.
The parameter $C_0$ is expected to be related to the difference in the
free energy per particle $\Delta F$ between a fluid particle in bulk
liquid, i.e., inside the three-dimensional reservoir, and one on the
surface of the substrate at $x=0$, and thus it is a measure of the
wettability of the substrate by the liquid \cite{Voue_lang}.
A general, explicit form for the relation between $C_0$ and $\Delta F$
is not available, but for a qualitative picture
\cite{Burlatsky_prl96, Voue_lang} an argument based on Langmuir-type
adsorption may be used to estimate
$C_0 \approx \sigma C_{reservoir} [1-\exp(-\beta \Delta F)]$,
where $\sigma$ is the area per adsorption site and $C_{reservoir}$ is
the density (number of particles per unit volume) in the reservoir.
At time $t=0$, the half-plane $x>0$ is empty.\\
%%%%%%%%%%%%%%%%%
\textbf{(b)} The substrate-fluid interaction is modeled as a periodic
potential forming a lattice of potential wells with coordination
number $z$ and lattice constant $a$. The particle motion proceeds via
activated jumps between nearest-neighbor wells; evaporation from the
substrate is not allowed. We assume that the dynamics of the activated
jumps can be described by the classical reaction-rate theory, i.e., the
activation barrier $U_A$ is significantly larger than the thermal
energy $k_B T$, where $k_B$ is the Boltzmann constant and $T$ the
temperature, and the coupling to the substrate is large enough such
that in crossing the barrier all the kinetic energy of the particle
is dissipated \cite{Hanggi}. The barrier $U_A$ determines the
jumping rate $\Omega = \nu_0 \exp[-U_A/k_B T]$, where $\nu_0$ is an
attempt frequency defining the time unit. We note that for a
two-dimensional homogeneous, isotropic substrate and a regular (square,
triangular, honeycomb, etc.) lattice structure, this jumping rate can
be absorbed into the one-particle diffusion coefficient
$D_0 = \Omega a^2/4$ on the bare substrate \cite{Lewis}. Therefore, in
this case $U_A$ is an irrelevant parameter in the sense that it can be
incorporated either into the choice of the unit of time as $\Omega^{-1}$
or into that of the unit of length as $\sqrt{D_0/\nu_0}$. For the rest of
this work we consider a square lattice ($z=4$); a qualitatively significant
dependence of the results on the lattice type is not expected. This
expectation is based on corresponding results obtained from test simulations
on a triangular lattice ($z=3$).\\
%%%%%%%%%%%%%%%%%%%%
\textbf{(c)} The pair interaction between fluid particles at distance
$r$ is taken to be hard-core repulsive at short range, preventing double
occupancy of the wells, and attractive at long range, $-U_0 /r^6$ for
$r \geq 1$, resembling a Lennard-Jones type interaction potential. Here
and in the following $r$ is measured in units of $a$ so that $U_0$
denotes the strength of the interaction energy. The absence of double
occupancy leads to an \textit{a priori} removal of thickening of the
film as a possible relaxation mechanism, which is not meant to imply
that we consider it irrelevant. We have decided to disregard this
mechanism here since it would have significantly increased the
complexity of the problem. Thus we leave the issue of film thickening
open for further research.\\
%%%%%%%%%%%%%%%%%%%%
\textbf{(d)} As we have mentioned in \textbf{(a)}, the motion proceeds
via activated jumps between nearest-neighbor wells, the activation
barrier for any jump being $U_A$. The selection of the nearest-neighbor
well, i.e., the probability $p(\bm{r} \to \bm{r'};t)$ that a jump from
location $\bm{r}$ will be directed toward the location  $\bm{r'}$, is
biased by the fluid-fluid energy landscape and is given by
\begin{equation}
p(\bm{r} \to \bm{r'};t) =
\frac
{
\exp \bigl\{ \frac{\beta}{2}
[\tilde U(\bm{r};t)-\tilde U(\bm{r'};t)]\bigr\}
}
{
Z(\bm{r};t)
},
\label{prob}
\end{equation}
where
$Z(\bm{r};t) =
\displaystyle{
\sum_{\bm{r'}, |\bm{r'} -\bm{r}|=1}
\exp \biggl\{ \frac{\beta}{2}
[\tilde U(\bm{r};t)-\tilde U(\bm{r'};t)]\biggr\}
}
$
is the normalization constant and $1/\beta~=~k_B T$,
\begin{equation}
\tilde U(\bm{r};t) = -U_0 \sum_{\bm{r'}, 0 < |\bm{r'} -\bm{r}|
\leq 3} \frac{\eta(\bm{r'};t)} {|\bm{r} -\bm{r'}|^6},
\label{potential}
\end{equation}
and $\eta(\bm{r'};t) \in \{0,1\}$ is the occupation number of the
well at $\bm{r'}$ at the time $t$. We note that, after canceling the
common factor $\exp[\beta \tilde U(\bm{r};t)/2]$, the expression
Eq.~(\ref{prob}) may be re-written in a form which is somewhat simpler
for the numerical simulations,
\begin{equation}
p(\bm{r} \to \bm{r'};t) =
\frac
{
\displaystyle{
\exp \left[ -\frac{\beta}{2}
\tilde U(\bm{r'};t)\right]
}
}
{
\displaystyle{
\sum_{\bm{r'}, |\bm{r'} -\bm{r}|=1}
}
\exp \left[ -\frac{\beta}{2}
\tilde U(\bm{r'};t)\right]
},
\label{prob_1}
\end{equation}
the dependence on $\bm{r}$ being retained because the summation is
carried out over the neighboring locations of $\bm{r}$. We also note
that the summation in Eq.~(\ref{potential}) has been restricted to
three lattice units for computational convenience. This corresponds
to the cut-off generally used in Molecular Dynamics simulations for
algebraically decaying Lennard-Jones pair-potentials.

The expression Eq.~(\ref{prob}) for the probability that a certain
direction is chosen for jumping deserves further discussion. For a
particle located at $\bm{r}$, it follows from the definition of
$p(\bm{r} \to \bm{r'};t)$ (Eq.~(\ref{prob})) that the rates
\begin{equation}
\omega_{\bm{r} \to \bm{r'};t} = \Omega p(\bm{r} \to \bm{r'};t)
\label{rate}
\end{equation}
for the transitions from $\bm{r}$ to neighboring locations $\bm{r'}$
satisfy
\begin{equation}
\sum_{\bm{r'}, |\bm{r'} -\bm{r}| = 1} \omega_{\bm{r} \to \bm{r'};t}
\equiv \Omega.
\label{total_rate}
\end{equation}
Thus the total rate of leaving a location for any given particle
at any given location is determined only by the fluid-solid
interaction characterized by $U_A$, it is time-independent, and
it equals $\Omega$. Therefore, the fluid-fluid interaction will
influence \textit{only} the choice of a \textit{direction} for the
jump, but not the jumping frequency, and in the dynamics there will
be only one relevant microscopic time scale, $\Omega^{-1}$, which is
dictated by the solid-fluid coupling.

The choice
$
p(\bm{r} \to \bm{r'};t) \propto \exp \bigl\{ \frac{\beta}{2}
[\tilde U(\bm{r};t)-\tilde U(\bm{r'};t)]\bigr\}
$
is motivated by the following. If one disregards the reservoir and
considers a system with a given volume and a given number of particles,
the change in the total fluid-fluid energy (no double occupancy)
$U_t = 1/2 \sum_{\bm{r}} \eta(\bm{r};t) \tilde U(\bm{r};t)$ due to
a change in configuration
\begin{equation}
\{\eta(\bm{r})=1,\eta(\bm{r'})=0\} \to \{\eta(\bm{r})=0,\eta(\bm{r'})=1\},
\label{jump}
\end{equation}
where $|\bm{r}-\bm{r'}| = 1$, is given by
$\Delta U_t = \tilde U(\bm{r'};t)-\tilde U(\bm{r};t)$, with
$\tilde U(\bm{r'};t)$ calculated for the final configuration and
$\tilde U(\bm{r};t)$ for the initial one.
Then a simple choice of transition rates $\omega'_{\bm{r} \to \bm{r'};t}$
which satisfies detailed balance with respect to the equilibrium
distribution
\begin{equation}
\mathcal{P} = \exp(-\beta U_t)/\mathcal{Z},
\label{gibbs}
\end{equation}
where $\mathcal{Z} = \displaystyle{
\sum_{all~\{\eta(\bm{r};t)\}} \exp(-\beta U_t)}$, is
\begin{equation}
\omega'_{\bm{r} \to \bm{r'};t} = \Omega \exp(-\beta \Delta U_t/2).
\label{rate_ff}
\end{equation}
If the transition rates would be chosen according to the expression
Eq.~(\ref{rate_ff}) above, then the fluid-fluid interaction would
effectively change the activation barrier and lead to a whole spectrum
of microscopic time scales. Normalizing these local transition rates
(Eq.~(\ref{rate_ff})) by the (local) total transition rate
$\displaystyle{\sum_{\bm{r'}, |\bm{r'} -\bm{r}| = 1}
\omega'_{\bm{r} \to \bm{r'};t}}$, a decoupling results between an
activated dynamics determined by the solid-fluid interaction and a
weak perturbation due to the fluid-fluid interaction, and
the choice given in Eq.~(\ref{prob}) for the bias probability
$p(\bm{r} \to \bm{r'};t)$ is obtained. One should note that this
decoupling is obtained at the expense that the transition rates
$\omega_{\bm{r} \to \bm{r'};t}$ defined by Eq.~(\ref{rate}) do not
satisfy detailed balance with respect to the distribution in
Eq.~(\ref{gibbs}), although the deviations, which are equal to
$Z(\bm{r};t)/Z(\bm{r'};t)$ due to
\begin{equation}
\frac{\omega_{\bm{r} \to \bm{r'};t}}{\omega_{\bm{r'} \to \bm{r};t}}
=\frac{Z(\bm{r};t)}{Z(\bm{r'};t)}\,\exp(-\beta \Delta U_t) ,
\label{deviations}
\end{equation}
are expected to be very small. From this point of view, the dynamics
defined by Eq.~(\ref{rate}) is that of an asymmetric exclusion
process \cite{Liggett}, but with a position- and time-dependent bias.
\\
%%%%%%%%%%%%%%%%%%%%%%
\textbf{(e)} As defined by the rules \textbf{(a)-(d)}, the model
corresponds to mass transport from the reservoir into a
two-dimensional vacuum so that a phase with very low-density, due
to two-dimensional evaporation, will form in front of the advancing
monolayer. The emergence of this low-density phase poses problems in
that its long time dynamics, which is of ideal gas type, mixes with
that of the following-up ``compact'' film and leads to serious
difficulties in defining the advancing edge of the monolayer. This
problem has been encountered earlier also in three-dimensional simulations
\cite{Kaski_92,Nissila_94,Koplik_96,DeConinck_96,Bekink_96,Voue_00}
and, in general, it has been overcome by replacing the simple particles
by connected chains mimicking polymers. Although this approach is
straightforward it is not very appealing, neither from a theoretical
point of view (an analytical approach becomes at least cumbersome,
if not intractable), nor from a computational one (the memory and CPU
requirements for sufficiently long simulation runs for large enough
systems are unreasonably high).

\begin{figure}
\centering
\includegraphics[width=.95\columnwidth]{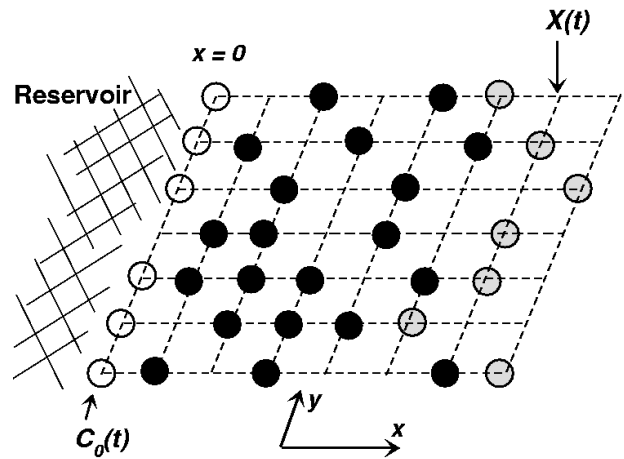}
\caption
{
\label{fig1}
Schematic drawing of a typical configuration $\{\eta(\bm{r};t)\}$ of
a monolayer spreading on a rectangular lattice (viewed under an oblique
angle). The particles denoted as open circles occupy the edge of the
reservoir $x=0$ with $C_0(t)$. Black circles denote the particles in
the bulk of the monolayer whereas gray circles denote particles at the
advancing edge $\Gamma_t$. Also indicated is the average position $X(t)$
of the advancing edge. Here $C_0(t)$ and $X(t)$ correspond to averages
over $y$ for a given realization and not to averages over runs (for which
we use the same notation in the main text). There are periodic boundary
conditions in the $y-$direction.
}
\end{figure}

Therefore, we have adopted a different approach. We define the
advancing edge $\Gamma_t$ of a monolayer configuration at time $t$ as
the set of the most advanced particles in each line $y = const$ for
this configuration (see Fig.~\ref{fig1}).
%%%%%%%%%%%%%%
%%%%%%%%%%%%%%
We eliminate two-dimensional ``evaporation'' by imposing the following
additional constraint: moves from sites $\bm{r} \in \Gamma_t$ toward
sites $\bm{r'}$ ahead of $\Gamma_t$ for which
$|\tilde U(\bm{r'};t)| < U_c$, where $U_c \geq 0$ is a fixed threshold
value, are rejected. This corresponds to requiring a given minimum
number of particles in the neighborhood $|\bm{r}| \leq r_c$ of any of the
components of $\Gamma_t$. The results presented in this paper correspond
to simulations with $U_c = U_0/3^6$, i.e., $r_c = 3$, in other words to the
requirement that in the disk $|\bm{r''} -\bm{r'}| \leq 3$ centered at
$\bm{r'}$ there is at least one more particle in addition to the one attempting
the jump $\bm{r} \to \bm{r'}$ (see also, c.f., Fig.~\ref{fig5}).

The above constraint is close in spirit to the ``effective
boundary-tension'' idea used in Ref.~\cite{Oshanin_jml} in which the
attractive interactions have been neglected except for particles on the
advancing edge for which a {\it constant} asymmetry in the jumping rates
``away'' and ``toward'' the reservoir was imposed. Rule \textbf{(e)}
provides a simple and convenient way of controlling the rate of
two-dimensional evaporation. For example, setting $U_c=0$ corresponds to
fully unconstrained dynamics, while replacing the rejection procedure with
an ``acceptance rate'' will allow for a continuous tuning of the evaporation
rate through the acceptance rate. We note that, physically, the model
defined by the rules \textbf{(a)-(e)} could be used to study also expansion
into an already present vapor phase instead of expansion into vacuum. In
the presence of a vapor phase there would be an average occupancy of the
sites in front of $\Gamma_t$, and thus some of the jumps from $\Gamma_t$
would be rejected due to the hard-core repulsion, which is an effect similar
to an acceptance rate as discussed above.\\
%%%%%%%%%%%%%%%%%%

\section{Kinetic Monte Carlo Simulations}

We have carried out KMC simulations of the model defined in Sect.~II
using square lattices with widths $L_y$ of 200 or 500 lattice units,
periodic boundary conditions along the transversal ($y$) direction
(appropriate for simulating an infinitely wide substrate), and an
activation energy $\beta U_A = 3.5$. Some simulation runs have been
carried out using lattices with smaller widths in order to check
finite-size effects. We have found that for widths larger than 100
lattice units there is no detectable influence of the width value on
the quantities we have measured in these simulations. The length of
the lattice in the $x$ direction has been chosen to be $L_x = 1000$
lattice units, with the possibility of changing it dynamically in the
course of the simulation  if necessary, i.e., if $\Gamma_t$ intersects
the line $x = 1000$; however, this situation was not encountered in
any of the simulations we have carried out. We note here that in the
experiments mentioned in Sect. I
\cite{Heslot_jpcm,Heslot_89,Heslot_prl,Daillant_90,Leiderer_92,
Fraysse_jcis,Voue_lang} typical values for the diffusion coefficient
were estimated from the spreading rate to be of the order of
$10^{-11} - 10^{-9} \, \textrm{m}^2/s$. If we take the lattice
spacing as $a \simeq 10 \, \textrm{nm}$, i.e., of the order of the
lateral size of a PDMS coil \cite{Perez}, then the above values for
the diffusion coefficient imply typical values for the frequency
$\Omega$ of the order of $10^5 - 10^7 \, \textrm s^{-1}$, and thus
for $\beta U_A = 3.5$ typical values for $\nu_0$ of the order of
$10^6 - 10^8 \, \textrm{s}^{-1}$.

For the simulations we have used a variable step continuous time
kinetic Monte Carlo algorithm \cite{Binder,Adam_KMC,Jansen} which
is described in Appendix A. One step in the Monte Carlo
simulation proceeds as follows. At time $t$, a particle from the film
($x \geq 0$) is selected at random. The time is incremented with
$\Delta t$ (the time at which a jump attempt with sufficient energy
for leaving the well \textit{will occur}), where $\Delta t$ is a random
variable distributed according to
$P(\Delta t)~=~N \Omega \exp(-N \Omega \Delta t)$
\cite{Binder, Adam_KMC,Jansen}, and $N$ is the number of particles in
the film at time $t$ (so that
$\displaystyle{\langle \Delta t\rangle_P = \frac{1}{N \Omega}}$).
The direction for the jump is chosen at random with probabilities
weighted according to Eq.~(\ref{prob}). If the destination site is
empty, the jump takes place; if not, the jump is rejected. Exchange
between the reservoir ($x<0$) and film ($x \geq 0$) is subject to
the additional constraint that the density
$\displaystyle{C_0(t) = \frac {1}{L_y} \sum_{y=1}^{L_y} \eta(x=0,y;t)}$
on the line $x=0$ fluctuates narrowly around a given value $C_0$ and
proceeds in the following manner. Moves from $x=0$ to $x = -1$ are
allowed if $C_0(t)$ is maintained within the interval
$[C_0 (1 - \delta),C_0 (1 + \delta)]$, where the amplitude $\delta$
has been fixed to $10^{-2}$. If this condition is satisfied, the
particle is considered to become part of the reservoir and is removed;
if not, the move is rejected. In case of moves from $x=0$ to $x = 1$,
if the density on $x=0$ would decrease below $C_0 (1 - \delta)$, then
after the move a new particle is added on an empty site (chosen at
random) on the line $x=0$. Similarly, in case of moves from $x=1$ to
$x = 0$ a particle is removed (at random) from the line $x=0$ if the
density would increase above $C_0 (1 + \delta)$. The time is not
incremented upon adding or removing particles, corresponding to the
reasonable assumption that the equilibration of the reservoir is very
fast. In order to compute the potential energy at a destination
site on the line $x = -1$, needed for the evaluation of the weight
probabilities for jumping of a particle on the line $x = 0$, in the
beginning of the simulation particles are placed at random on the
lines $-4 \leq x \leq -1$ such that the average density on these lines
is $C_0$, and this configuration is kept unchanged during the simulation
run. Due to these procedures one does not have to consider the dynamics
on the lines $x \leq -1$, i.e., in the reservoir. In order to have
sufficient fluctuations in $C_0(t)$ on the line $x = 0$, fluctuations
which mimic the stochastic nature of the exchange of particles between
the reservoir and the film, the width $L_y$ has to be large enough
such that the amplitude $\delta$ translates into a reasonable number
of sites. This is the reason why we have used a large value for the
width; for example, at $C_0 = 0.8$ and $L_y = 500$, $\delta=10^{-2}$
translates into $4$ sites.

All the ``measured'' quantities have been averaged over a number of
independent simulation runs ranging from 10 to 50, a value of 50
runs being used in most of the cases. These runs differ from each
other both with respect to the initial configuration
$\{\eta(x=0,y;t=0)\}$ and the subsequent sequence of jumps. The
observables of interest are defined below. The density $\rho(\bm{r};t)$
is defined as $\rho(\bm{r};t)=\left\langle\eta(\bm{r};t)\right\rangle$,
where $\langle \cdots \rangle$ means average over different KMC runs.
Due to the symmetry of the model, the density profile
$\tilde\rho(\bm{r};t)$ in the limit of infinitely many runs is
independent of $y$, while in the average over a finite number of
runs random, uncorrelated fluctuations (whose amplitude decrease
with increasing number of runs) occur along the $y$ direction.
These fluctuations are suppressed by measuring the transversally averaged
density $\displaystyle{C(x,t) = \left\langle \frac{1}{L_y}
\sum_{y=1}^{L_y} \eta(x,y;t) \right\rangle}$,
and thus it is expected that $C(x,t) \simeq \tilde\rho(\bm{r};t)$,
with strict equality for infinitely many runs. The average position
of the advancing edge of the monolayer is defined as
$\displaystyle{X(t) = \left\langle \frac{1}{L_y}
\sum_{\bm{r} \in \Gamma_t} x\right\rangle}$.
For the case $U_c = U_0/3^6$ (as used for the actual simulations),
two-dimensional evaporation is negligible and $X(t)$ (which we shall
also call front line) is a good measure for the actual advancing edge
of the monolayer. We note here that in the case when no constraint is
imposed to prevent two-dimensional evaporation ($U_c = 0$), the front
line may be defined as $X(t) = \langle\hat x(t)\rangle$ \cite{Neogi_jcp},
where $\hat x(t)$ is the most advanced line corresponding to a given
(smallest measurable) density
$\displaystyle{\hat C = \frac{1}{L_y} \sum_{y=1}^{L} \eta(\hat x,y;t)}$.
Alternatively, one may follow the time dependence of the mass of the
film \cite{Lacasta}, or use a percolation type definition for the precursor
as the set composed of all the particles which are connected along
nearest-neighbor bonds with the reservoir, the boundary of this cluster
defining $\Gamma_t$~\cite{Abraham_02}.

\begin{figure*}[!htb]
\begin{minipage}[c]{.9\textwidth}
\par{\hspace*{1.in} $t = 2 \times 10^5$ \hspace*{1.4in} $t = 10^6$
\hspace*{1.4in} $t = 2 \times 10^6$\hfill}
\vskip .1in
\includegraphics[width=.98\textwidth]{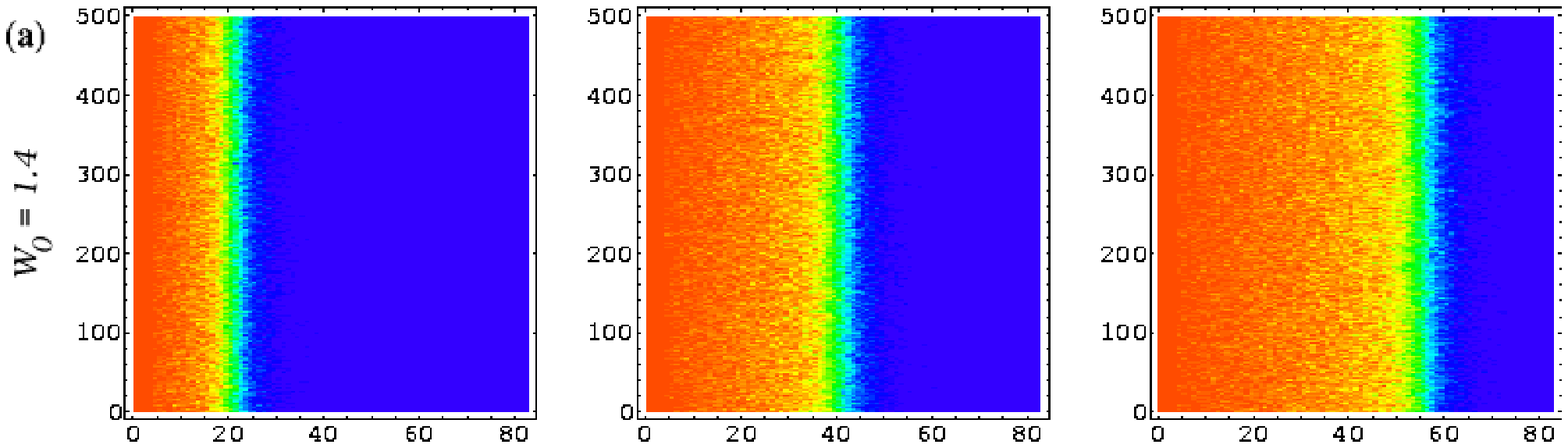}
\vskip .1in
\includegraphics[width=.99\textwidth]{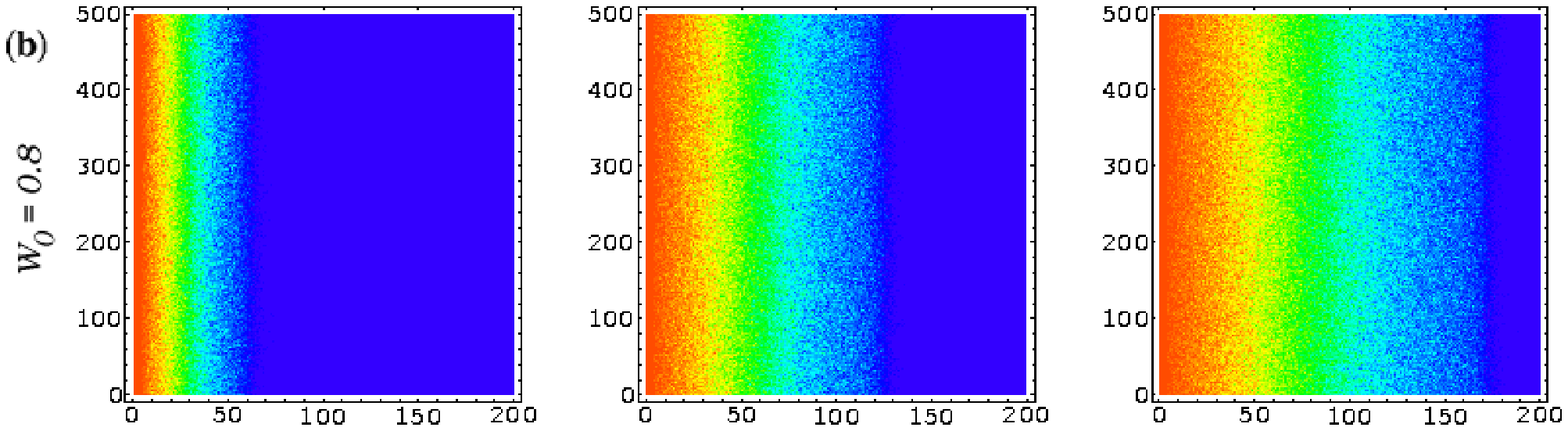}
\end{minipage}%
\begin{minipage}[c]{.1\textwidth}
\vskip .5in
\includegraphics[width=.9\textwidth,height=5.cm]{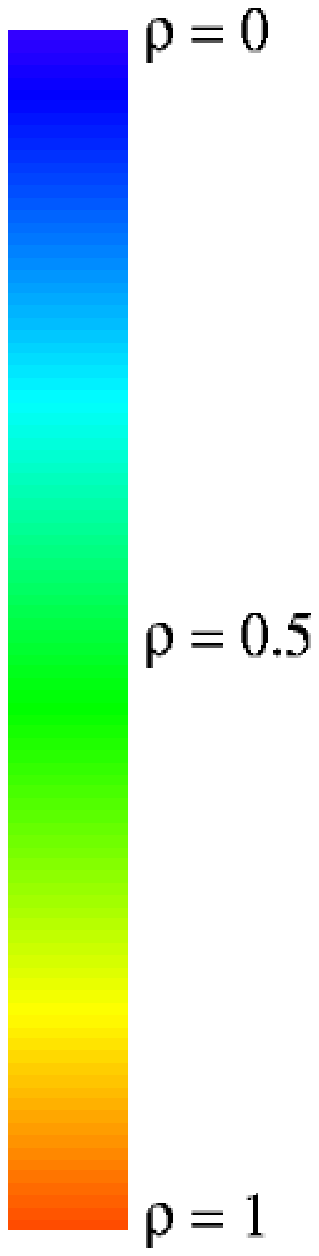}
\vskip .3in
\includegraphics[width=.9\textwidth]{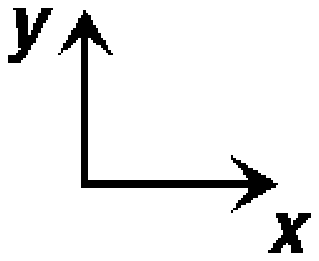}
\end{minipage}
\caption
{
\label{fig2}
Typical density profiles $\rho(x,y;t)$ (obtained by averaging over
50 KMC runs) for $W_0 = \beta U_0 = 1.4$ (row (a)) and
$W_0 =0.8$ (row (b)) at times (left to right)
$t = 2 \times 10^5$, $t = 10^6$, and $t = 2 \times 10^6$ for
$C_0 =1.0$ and $L_y = 500$. Time is measured in units of
$\nu_0^{-1} = \Omega^{-1} \exp(-\beta U_A)$ with $\beta U_A = 3.5$,
distances are measured in lattice units, and spreading occurs in
$x-$direction. The color coding (shown on the right) is a linear
function of density.
}
\end{figure*}

Snapshots of typical density profiles during spreading are shown in
Fig.~\ref{fig2} for the cases (a) $W_0 = 1.4$ and (b) $W_0 = 0.8$,
where we have introduced the notation
%%%%%%%%%%%%%%
\begin{equation}
W_0 = \beta U_0.
\label{def_U0}
\end{equation}
%%%%%%%%%%%%%%

%%%%%%%%%%%%%%
%%%%%%%%%%%%%%
These density profiles reveal already the qualitative dynamical
behavior. It can be seen that, as expected, the monolayer is
homogeneous in the $y$ (transversal) direction, while along the
$x$ (spreading) direction there are significant density variations.
As intuitively expected, the spreading of the \textit{edge} of the
monolayer is faster for smaller $W_0$, i.e., higher temperature
(at a given interaction strength $U_0$) or smaller inter-particle
attraction (at a given temperature $T$). In addition, one observes
qualitatively different dynamics as revealed by the abrupt change
from high to low density for the large value of $W_0$ compared to
the smooth and broad decrease for the small value of $W_0$. This is
accompanied by a different dynamics of the regions with moderate
to low density in the two cases. While for large value of $W_0$
(panel (a)) the range of densities $\rho \simeq 0.5$ (the green band)
and that of low densities $\rho \lesssim 0.2$ (the light blue band)
maintain relatively constant widths and advance with similar rates,
for smaller $W_0$ (panel (b)) the regions become broader with
increasing time and clearly the low density (light blue) has a
larger rate of advancing. These features will be discussed more
in the following sections, where a quantitative characterization of
the spreading as a function of the parameters $W_0$ and $C_0$ is
presented.

\section{Time-Dependence of Spreading
and Asymptotic Scaling of Densities}

\subsection{Time dependence of spreading}
\begin{figure}[htb!]
\begin{minipage}[c]{.95\columnwidth}
\includegraphics[width=.95\textwidth]{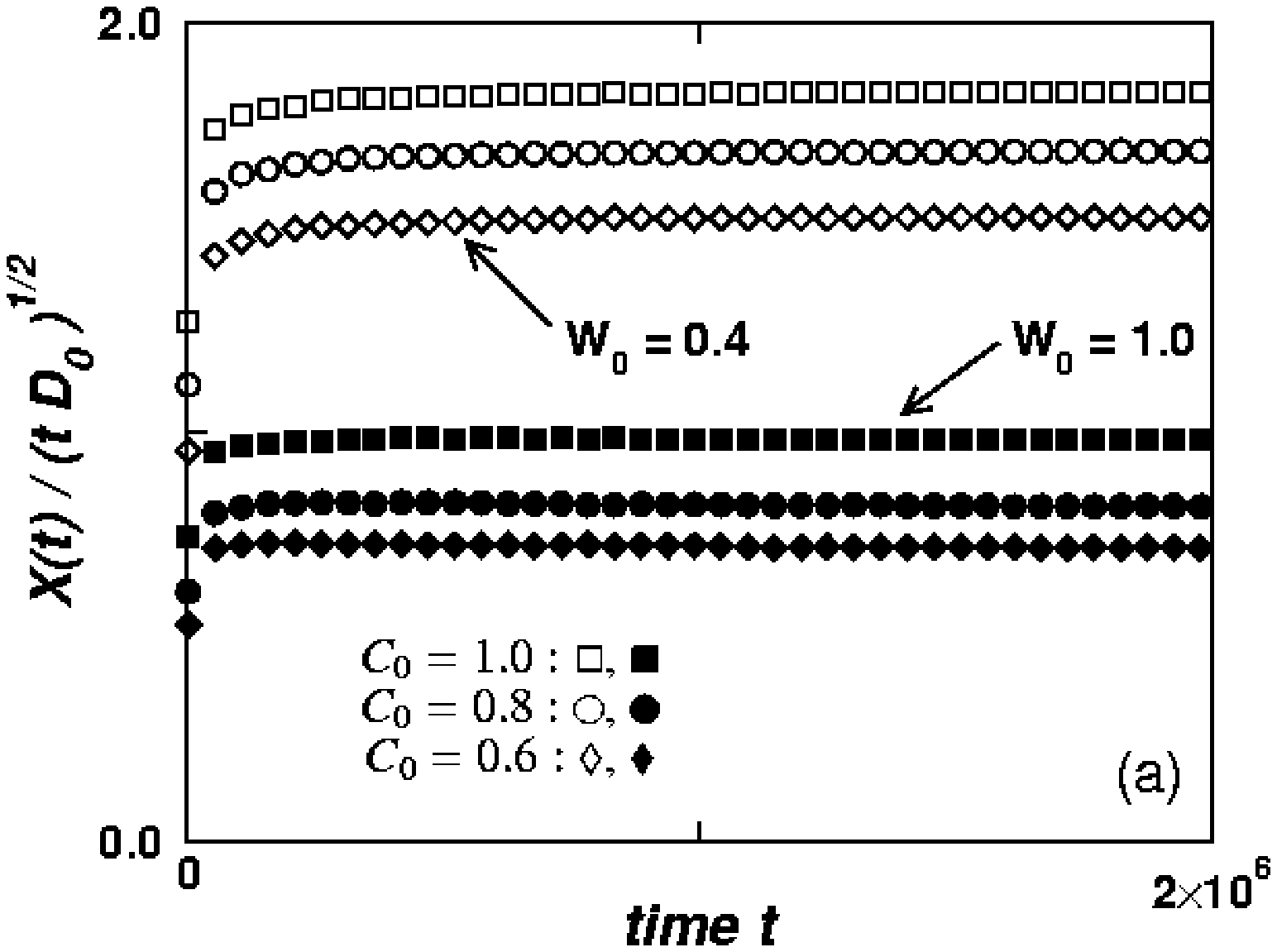}%
\end{minipage}%

\begin{minipage}[c]{.95\columnwidth}
\hspace*{-.2in}%
\includegraphics[width=.95\textwidth]{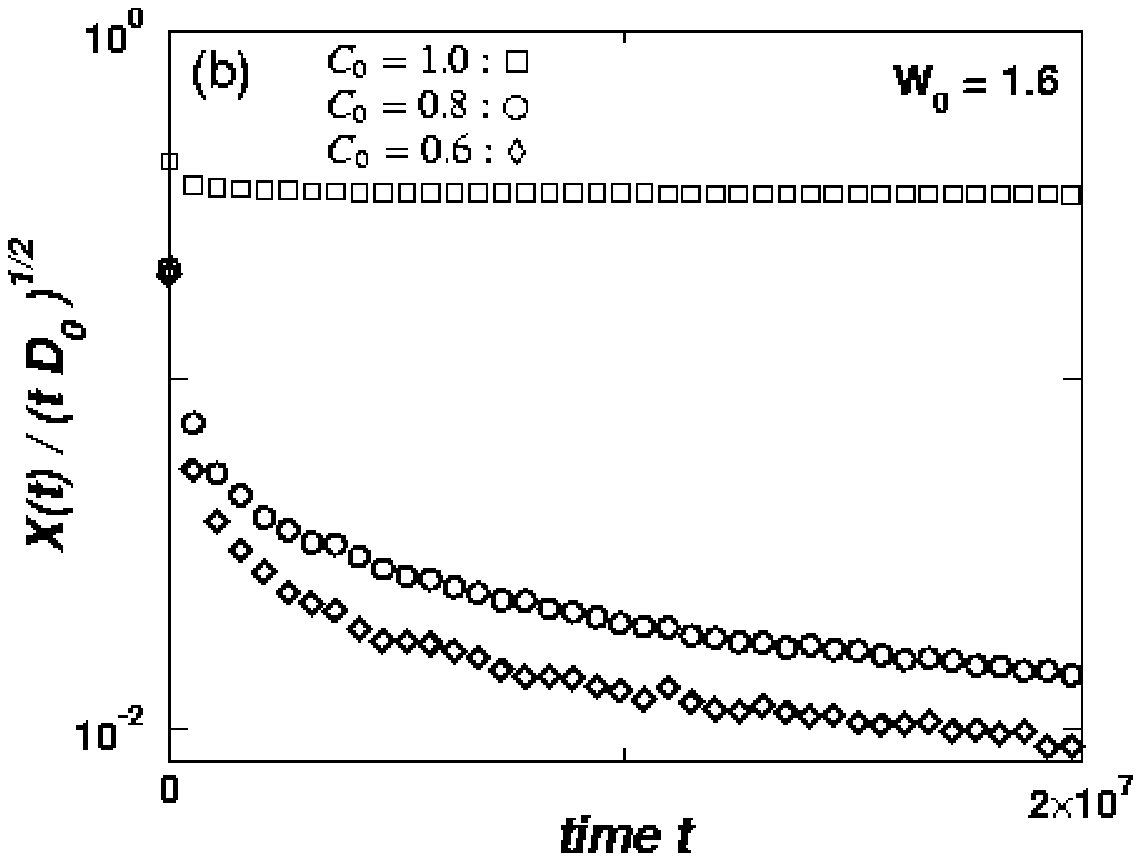}%
\raisebox{.210\textwidth}
{
\hspace*{-1.55in}%
\includegraphics[width=.4\textwidth]{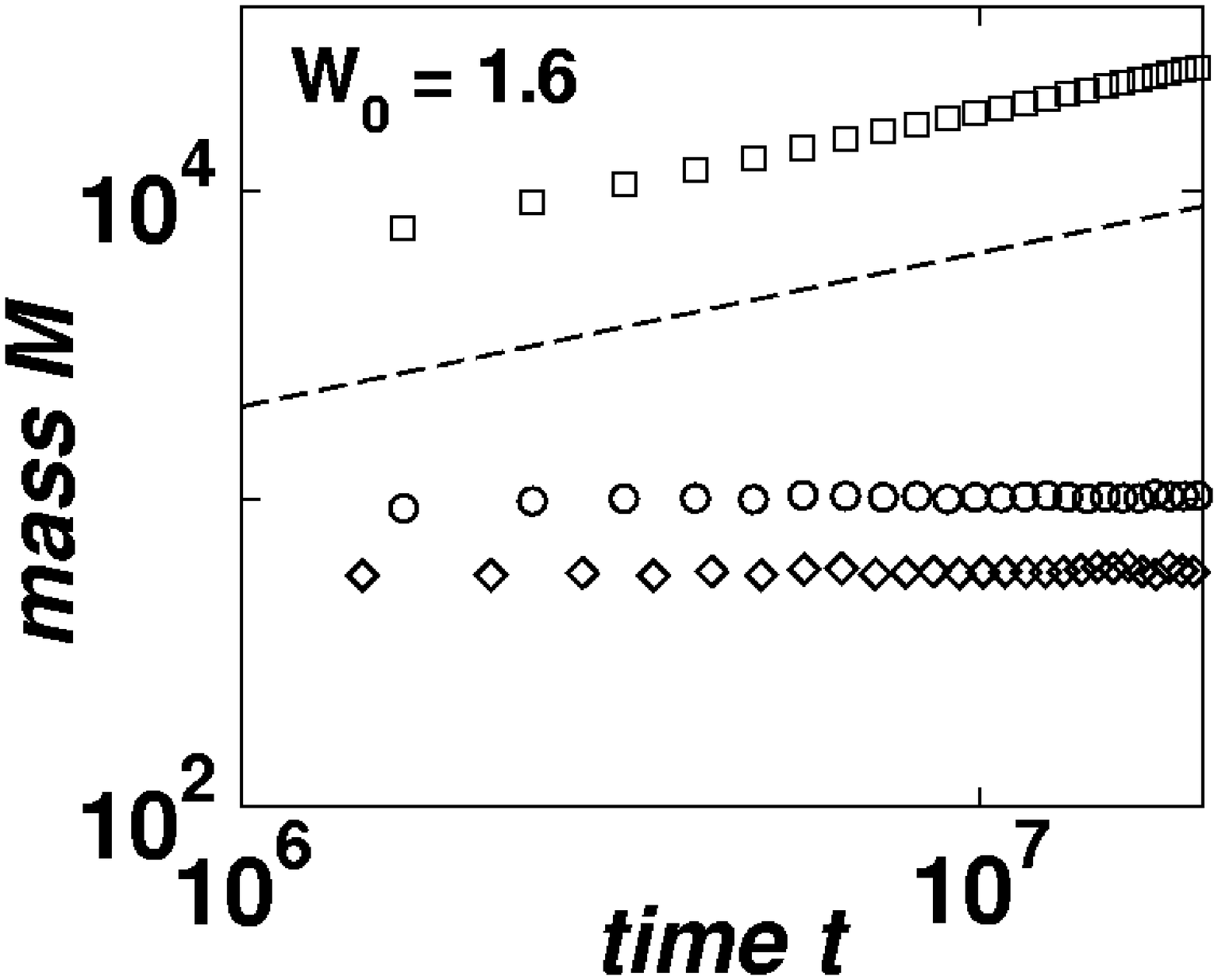}%
}%
\end{minipage}%
\caption
{
\label{fig3}
Front line position (divided by $\sqrt{D_0 t}$) as a function of time
for (a) $W_0 = 0.4$ (open symbols) and $W_0 = 1.0$ (filled symbols),
and (b) $W_0 = 1.6$, for reservoir densities $C_0=1.0$ (squares),
$C_0=0.8$ (circles), and $C_0=0.6$ (diamonds).
The inset (on logarithmic scales) compares the time dependence of the
(unscaled) mass $M(t)$ of the film for $W_0 = 1.6$ and for the same
reservoir densities (symbols) with the $t^{1/2}$ behavior (dashed line).
In (b) for $C_0 = 0.8$ and $C_0 = 0.6$ the scaled front line position
decays $\sim t^{-1/2}$ for large $t$. Here and in the following time
$t$ is measured in units of $\nu_0^{-1} = \Omega^{-1} \exp(-\beta U_A)$
with $\beta U_A = 3.5$, so that
$\sqrt{D_0 t} = \sqrt{\nu_0 t} \,\,\frac{a}{2} \,
\exp(-\frac{1}{2}\beta U_A)$.
}
\end{figure}

We start our analysis of the dynamics of spreading by studying the
time-dependence of the position $X(t)$ of the front line. As noted in
the Introduction, we expect that the asymmetry in the jumping rates
due to the inter-particle interactions (Eq.~(\ref{prob})) will not
affect the previously (without inter-particle interactions) predicted
$X(t) \sim \sqrt t$ time dependence as long as the attractive
interaction is not too strong.
Figure~\ref{fig3} shows the time dependence of the scaled front line,
$X(t)/\sqrt{D_0 t}$, for several values of the interaction parameter
$W_0$ and of the reservoir density $C_0$.

%%%%%%%%%%%%%%
%%%%%%%%%%%%%%

One can see that at low and intermediate values of the
interaction strength (Fig.~\ref{fig3}(a)) the $\sqrt{t}$ behavior
is recovered independent of the value $C_0$ of the density in the
reservoir. However, for strong attractive fluid-fluid interaction
(Fig.~\ref{fig3}(b)) and low densities $C_0$, the time dependence
of $X(t)$ clearly deviates from the $\sqrt{t}$ behavior, the latter
being obtained only for high densities $C_0$. Since the decreasing
trend shown by the data in Fig.~\ref{fig3}(b) may be due to either
spreading at a rate slower than $\sqrt{t}$, or to the fact that there
is no extraction of a macroscopic film from the reservoir, we have
also analyzed the time dependence of the total mass of the film
extracted, $\displaystyle{ M(t)=
\left\langle\sum_{x > 0,\, y}\eta(\bm{r};t)\right\rangle}$.
As can be seen in the inset in Fig.~\ref{fig3}(b), for large $C_0$
the time dependence of $M(t)$ is very well described by $\sqrt{t}$,
as expected \cite{Burlatsky_prl96}, while for low values $C_0$ the mass
$M(t)$ shows a clear saturation to a small constant value (which
corresponds roughly to a position of the front of ca. 10 lattice
units), thus indicating that actually there is no macroscopic film
extracted from the reservoir. At early times fluctuations lead to
the extraction or leakage (with a very fast decreasing rate
of extraction) of a small number of particles from the reservoir,
but the spatial extension $X(t)$ of the film in this case remains
microscopicly small and becomes time independent for $t \gg 1$.
Therefore, as a function of the interaction strength $W_0$ there
is a transition from a ``substrate covering'' state at low values
$W_0$, in the sense of extraction of a film with macroscopic lateral
extension in the spreading direction independently of the density
value $C_0$ in the reservoir edge, i.e., a film which spreads
according to $X(t \to \infty) \sim \sqrt{t}$, to a ``non-covering''
state at large values $W_0$, in the sense that a macroscopic film
is extracted only for sufficiently large densities $C_0$ (eventually
for none if $W_0$ is sufficiently large). Results similar to the
ones shown in Fig.~\ref{fig3}, from simulations performed for a
broad range of parameters values ($0.4 \leq W_0 \leq 1.6$ and
$0.1 \leq C_0 \leq 1.0$) indicate that the values $W_0^{(cov)}(C_0)$
for which this change in behavior occurs are bounded from below by
$1.0 < W_0^{(cov)}(C_0)$. We shall return to this point in the next
but one paragraph and during the discussion of the continuum limit.

\begin{figure}[htb!]
\begin{minipage}[c]{.45\textwidth}
\hspace*{-1.8in}%
\includegraphics[width=.95\textwidth]{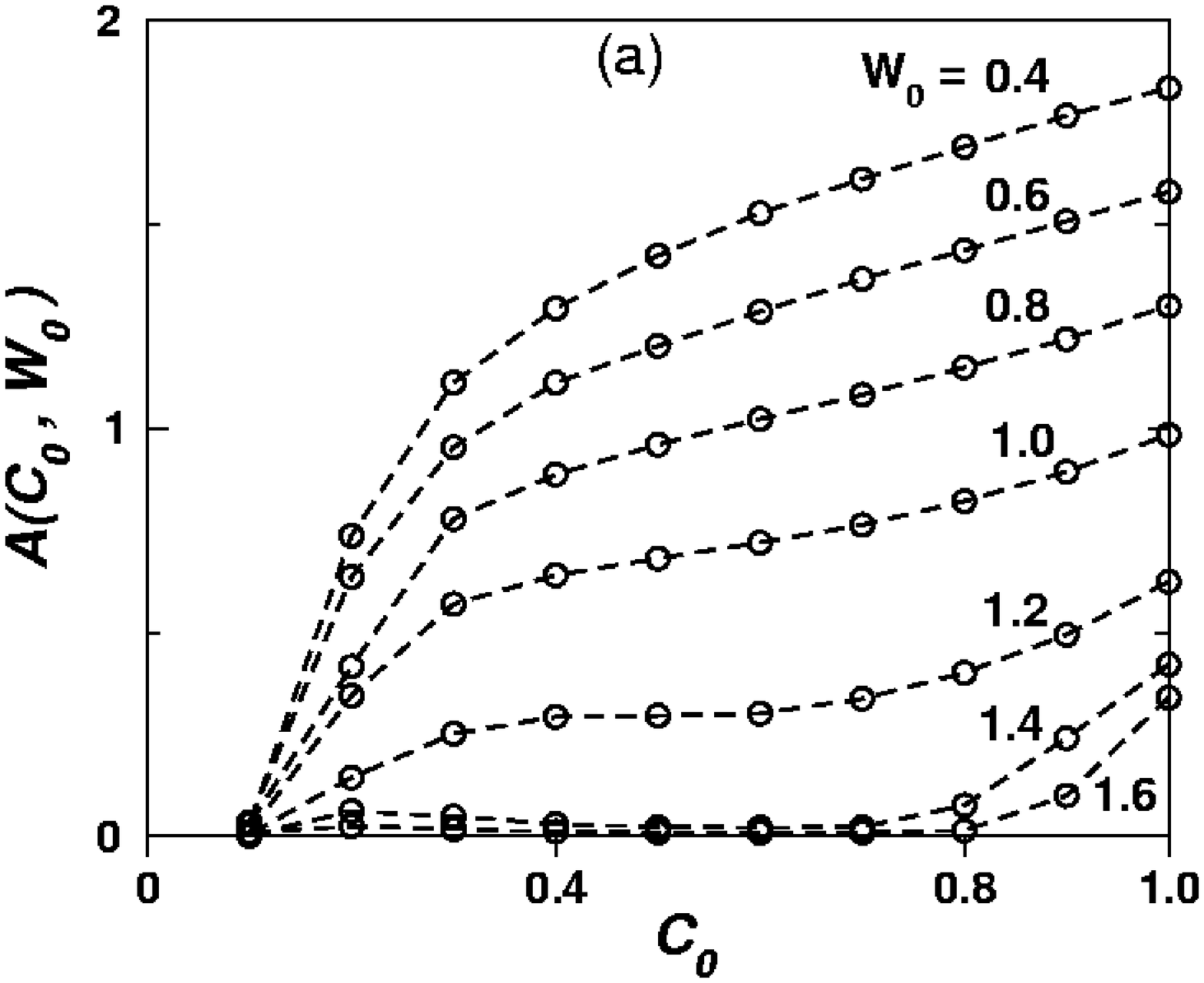}%
\raisebox{.512\textwidth}
{
\hspace*{-2.65in}%
\includegraphics[width=.27\textwidth]{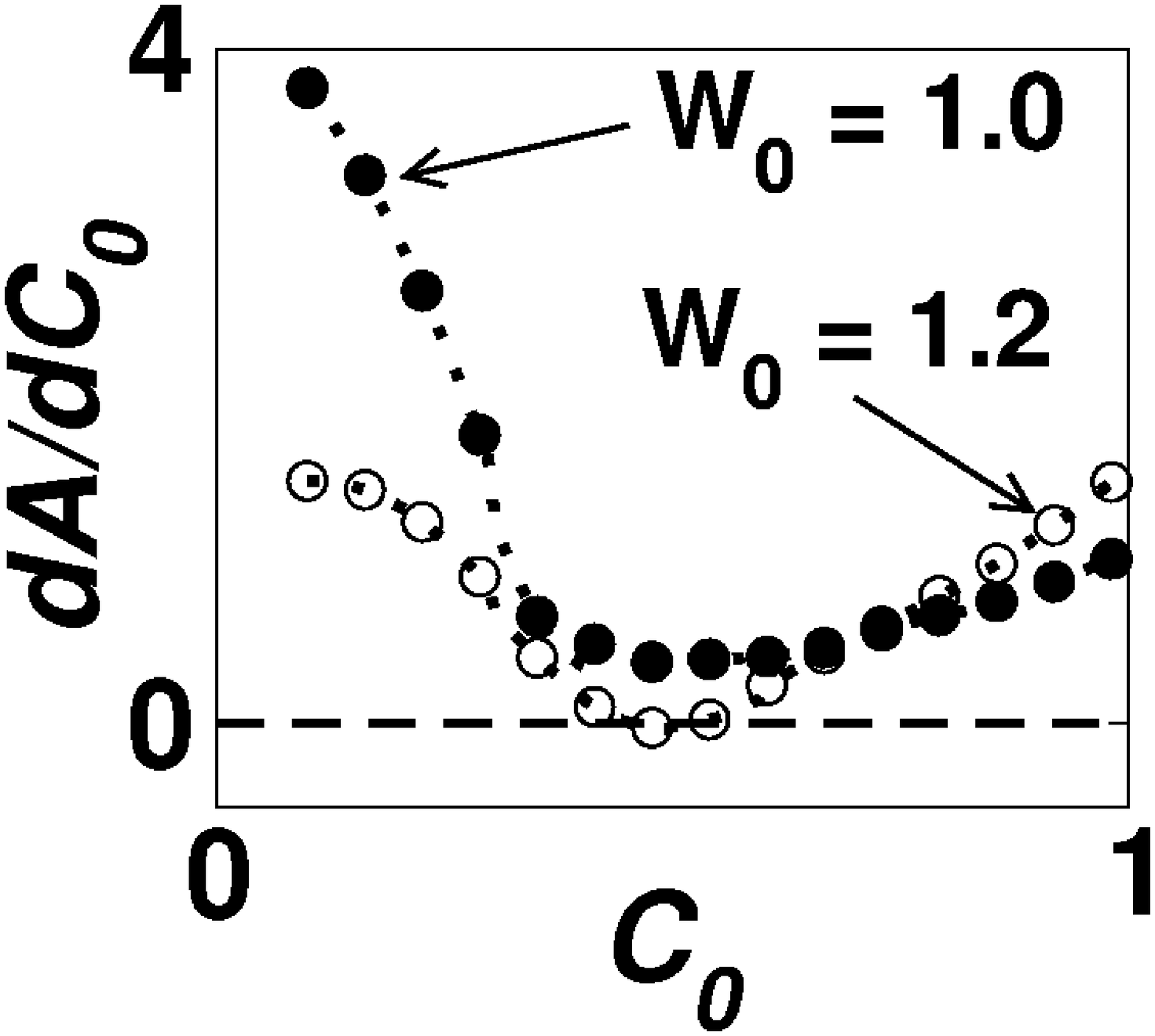}%
}%
\end{minipage}%

\begin{minipage}[c]{.45\textwidth}
\hspace*{-.45in}%
\includegraphics[width=.98\textwidth]{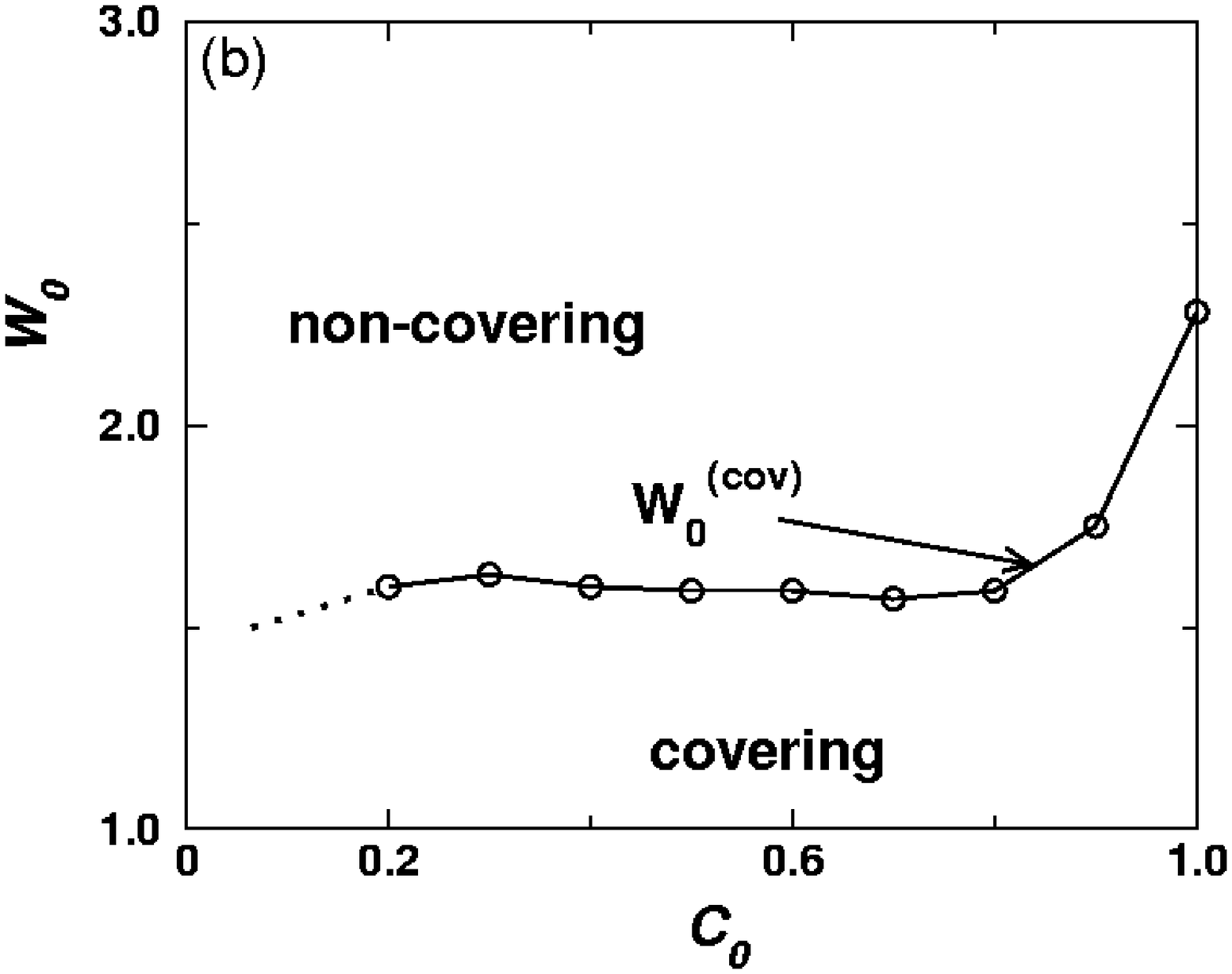}%
\raisebox{.44\textwidth}
{
\hspace*{-1.4in}%
\includegraphics[width=.35\textwidth]{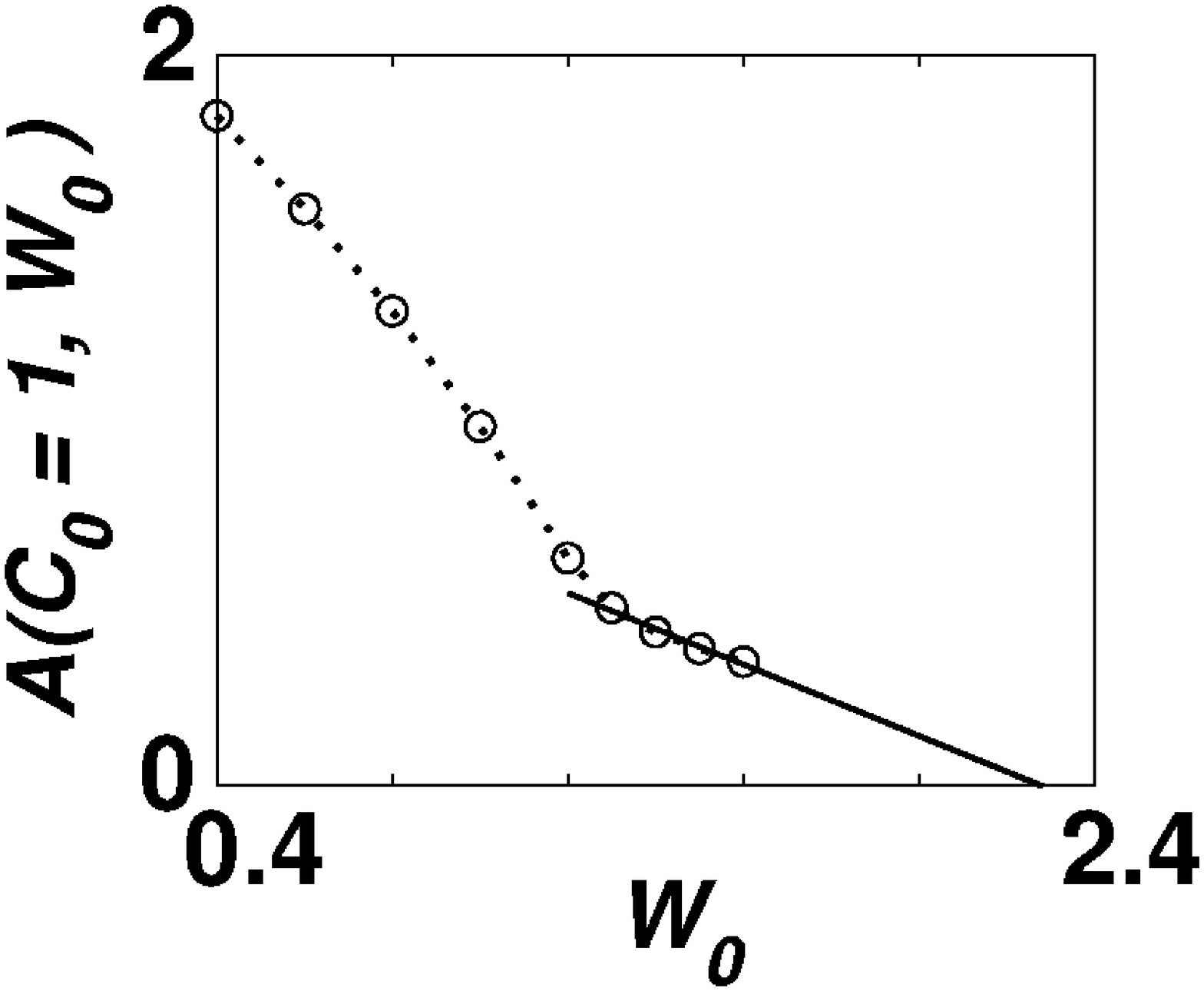}%
}%
\end{minipage}%
\caption
{
\label{fig4}
(a) Dependence of the prefactor $A(C_0,W_0)$ on $C_0$ for
several values of $W_0$. The inset shows the derivative
$dA(C_0)/dC_0$ as a function of $C_0$ in the cases
$W_0 = 1.2$ ($\Circle$) and $W_0 = 1.0$ ($\CIRCLE$);
the dashed line in the inset corresponds to $dA(C_0)/dC_0 = 0$.
(b) Estimate of a ``covering phase-diagram'' in the
$W_0 \textrm{--} C_0$ plane showing the parameter range for
which a macroscopic film is extracted from the reservoir
or is not extracted, respectively. The curve $W_0^{(cov)}$
shows the ``covering -- non-covering'' separatrix described
in the main text. The line is a guide for the eye, and the
dotted line at $C_0 \leq 0.2$ is a heuristic extrapolation
suggesting $W_0^{(cov)}(C_0 \to 0) \simeq 1.4$.
The inset shows the dependence of $A(C_0 = 1, W_0)$ on
$W_0$. The line extrapolated to $A(C_0 = 1, W_0) = 0$ is
a linear fit of the last four points. In all four plots the
symbols are KMC results; dashed and dotted lines connecting
the symbols are guides to the eye.
}
\end{figure}

The results presented in Fig.~\ref{fig3} also show that in case of
spreading the time independent dimensionless prefactor $A$ in
$X(t \to \infty) = A \sqrt{D_0 t}$ depends on both $W_0$ and $C_0$.
From the curves $X(t)/\sqrt{D_0 t}$ one can estimate
$A=\displaystyle{\lim_{t \to \infty} X(t)/\sqrt{D_0 t}}$ by fitting the
data in the range $t \gg 1$ (in practice the data in the last tenth
of the time interval available) with a constant. The results for
$A(C_0,W_0)$ are shown in Fig.~\ref{fig4}(a).
%%%%%%%%%%%%%%
%%%%%%%%%%%%%%
Here we use the convention that in the case where there is no
macroscopic film spreading in the sense explained above, i.e.,
$X(t)/\sqrt{D_0 t}$ decreases in time and $M(t)$ has a time
dependence $t^{\alpha_1}$ with $\alpha_1$ significantly smaller
than~$1/2$ (in practice $\alpha_1 < 0.4$), the value of the
prefactor is assigned to be $A = 0$. As expected, $A(C_0,W_0)$
is an increasing function of $C_0$ for fixed $W_0$ and a
decreasing function of $W_0$ for fixed $C_0$.
However, the functional dependence is not simple, and one can
easily notice a change in the shape of $A(C_0,W_0)$
for $W_0$ close to the value $1.0$. For values
$W_0 \gtrsim 1$, the curve shows a plateau over a range of
densities $C_0$ which increases with increasing $W_0$, while
for values $W_0 \lesssim 1$ the prefactor $A$ is a strictly
increasing function of $C_0$. This property is of a different
type than the transition from covering to non-covering discussed
above, because it involves a change in the dependence of the
prefactor $A$ on the density $C_0$ while the spreading law
$X(t) \sim t^{1/2}$ holds. This change in behavior emerges
(as we shall discuss in more detail in the next section) as a
consequence of the competition between the diffusive motion
driven by the concentration gradient and the clustering
tendency (opposing the concentration gradients) driven by the
inter-particle attraction. This competition leads to
instabilities at certain density values if the attractive
interaction is sufficiently strong. From Fig.~\ref{fig4}(a),
the value of the threshold interaction $W_0^{(t)}$ for the
onset of a plateau can be estimated to be bounded as
$1.0 < W_0^{(t)} < 1.2$. As shown in the inset, these bounds
are confirmed also by the behavior of the derivative
$dA(C_0)/dC_0$, which is clearly larger than zero for
$W_0 = 1.0$, and becomes zero (within the limits of numerical
accuracy) around $C_0 \simeq 0.5$ for $W_0 = 1.2$.

For $C_0$ fixed the values of the prefactor $A(C_0,W_0)$
as a function of $W_0$ can be used to estimate via linear
extrapolation, as shown in the inset in Fig.~\ref{fig4}(b)
for the particular value $C_0 = 1$, the interaction strength
$W_0^{(cov)}(C_0)$ at which for a given $C_0$ the prefactor
vanishes: $A(C_0, W_0^{(cov)}) = 0$. The data for
$W_0^{(cov)}(C_0)$ are shown in Fig.~\ref{fig4}(b). Since at
a given $W_0$ the prefactor attains its maximum for $C_0 = 1$,
the dependence of $A(C_0=1, W_0)$ on $W_0$ allows one to infer
also an estimate of the interaction $\widetilde W_0^{(cov)}$
above which no macroscopic film is extracted from the reservoir
whatever the density $C_0$ at the reservoir edge is, i.e.,
$A(C_0, W_0 > \widetilde W_0^{(cov)}) \equiv 0$. This value
cannot be measured directly due to the unreasonably long
simulation times needed to reach the asymptotic regime in this
range of $W_0$ values, but the linear extrapolation of the
available data, as shown in the inset of Fig.~\ref{fig4}(b),
yields the estimate $\widetilde W_0^{(cov)} \simeq 2.3$.

\begin{figure}[!htb]
\centering
\includegraphics[width=.6\columnwidth]{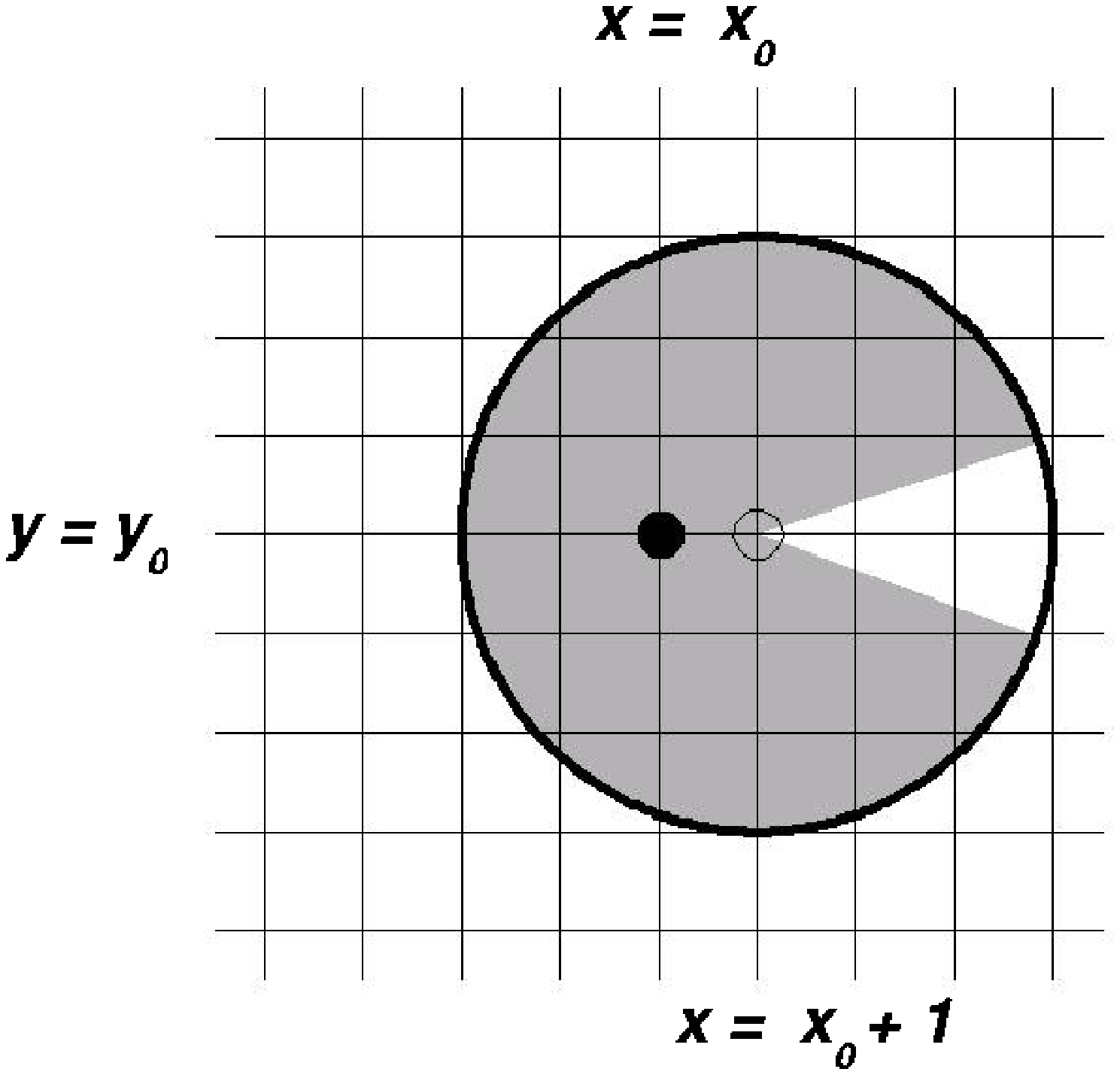}
\includegraphics[width=.2\columnwidth]{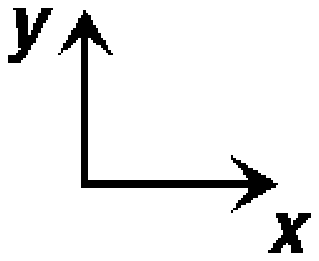}
\caption
{
\label{fig5}
Schematic drawing of the region around a point $(x_0,y_0)$
(filled circle) belonging to the \textit{advancing} edge $\Gamma_t$.
The target destination, for the case of an \textit{advancing} $\Gamma_t$,
is denoted by an empty circle. The shaded area shows the domain in
which at least one other site must be occupied so that the particle at
$(x_0,y_0)$ can move to $(x_0+1,y_0)$ in accordance with our KMC rule
\textbf{(e)} (see Sect. II). Since $(x_0,y_0) \in \Gamma_t$, all
sites $(x>x_0,y_0)$ must be empty so that in the unshaded sector there
are no occupied sites.
}
\end{figure}

In the range of low densities $C_0$ at the reservoir edge, the KMC
results in Fig.~\ref{fig4} indicate that there is a threshold value
$C_0^{(min)} \simeq 0.1$ for the density in the reservoir edge below
which, independent of the interaction strength $W_0$, there is no
extraction of a monolayer: all the curves $A(C_0)$ reach zero at a
non-zero value of $C_0$. As we will show below, this is a consequence
of the condition \textbf{(e)} in the model, i.e., of the requirement
that a move from $\bm{r} \in \Gamma_t$ toward a forward site $\bm{r'}$
is accepted only if there is at least another particle in the neighborhood
$|{\bm{r}}'-\bm{r}| \leq r_c$. As we have mentioned before, this
condition is equivalent to requiring a minimum value $C_1$ for the
density on the advancing edge of the film. This density $C_1$ can be
estimated as follows. For a fluid particle at $(x_0,y_0) \in \Gamma_t$
(the filled circle in Fig.~\ref{fig5}) to move to the site $(x_0+1,y_0)$
(the open circle in Fig.~\ref{fig5}), in the shaded area of the disk
of radius $r_c = 3$ centered at $(x_0+1,y_0)$ there must be at least one
more particle in addition to the one attempting the jump. The unshaded
sector of the disk is due to the fact that by definition of $\Gamma_t$
the sites $(x > x_0,y_0)$ are empty.
%%%%%%%%%%%%%%
%%%%%%%%%%%%%%
Therefore, at least two of the $M=25$ sites in this shaded region are
occupied, and thus $C_1 = 0.08$ is an estimate for the minimum density
on $\Gamma_t$. Since extraction of a film requires a \textit{moving}
edge $\Gamma_t$, the density at the reservoir edge should be greater
than $C_1$ for spreading to be possible. This explains the threshold
value $C_0^{(min)} \simeq 0.1$ observed in the simulations.

\subsection{Asymptotic scaling}

\begin{widetext}
%%%%%%%%%%%%%%
\begin{figure}[tbh!]
\begin{minipage}[c]{.5\textwidth}
\hspace{-.25in}%
\includegraphics[width=.95\textwidth]{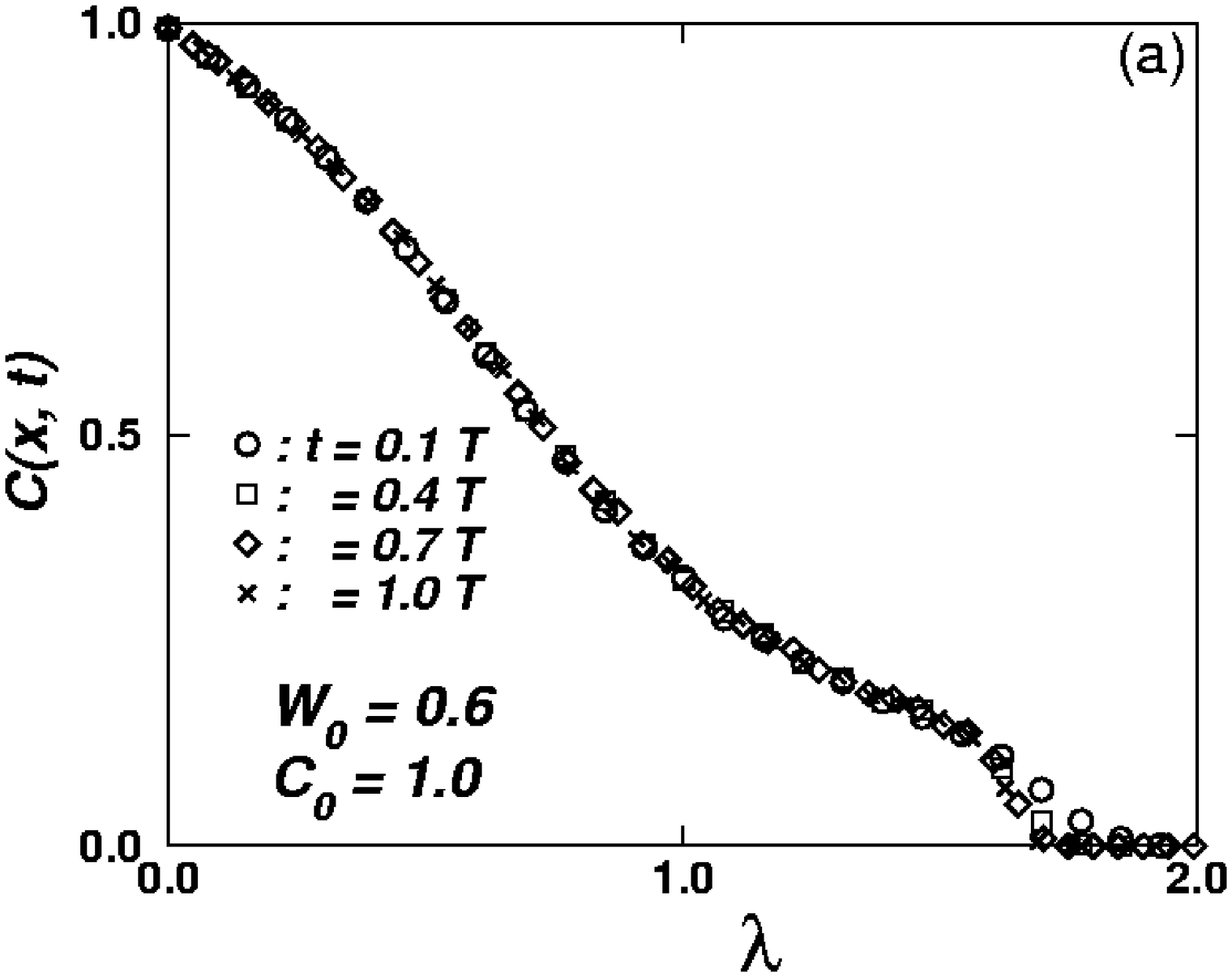}%
\raisebox{.33\textwidth}
{
\hspace{-1.6in}%
\includegraphics[width=.4\textwidth]{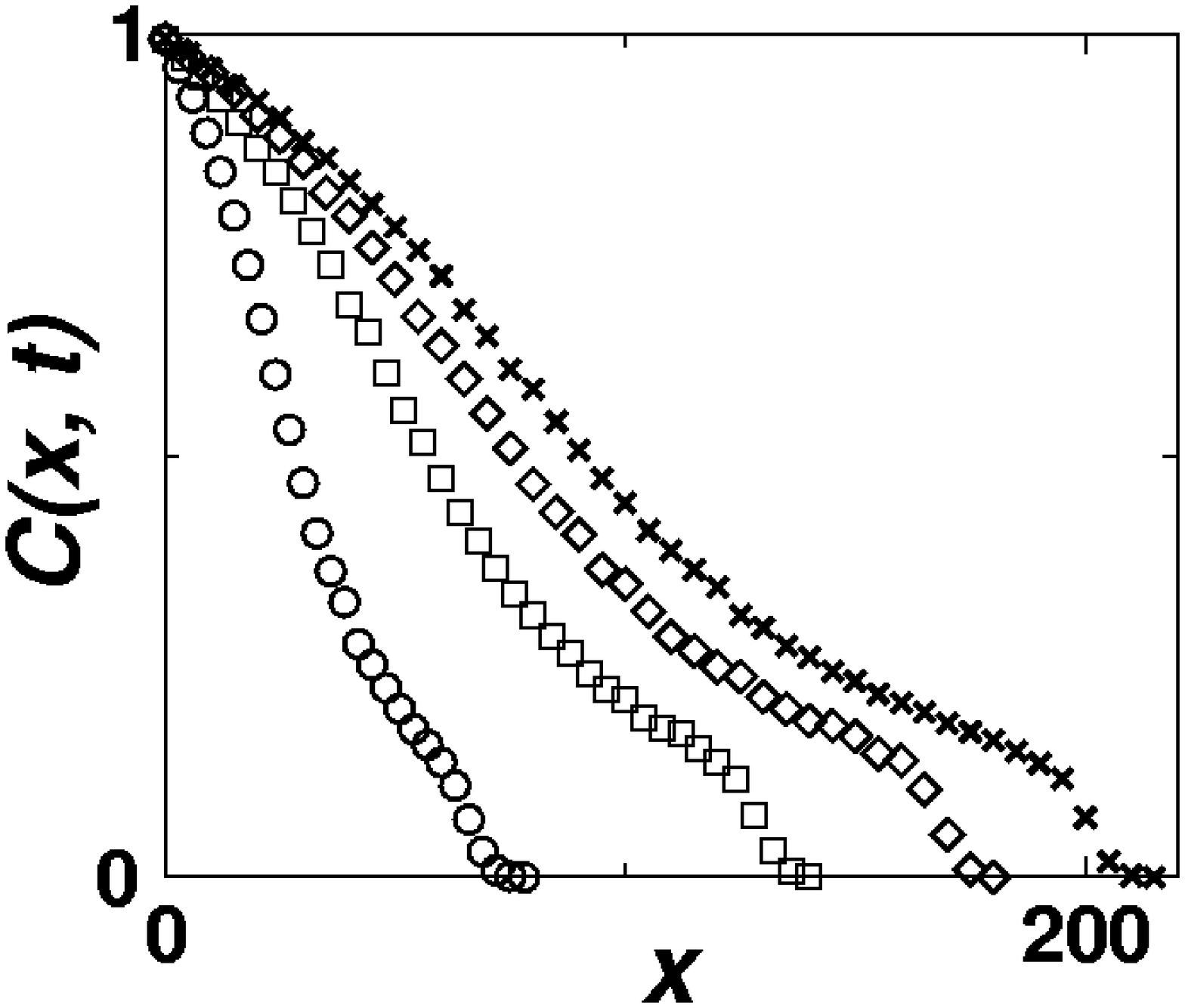}%
}%
\end{minipage}%
\begin{minipage}[c]{.5\textwidth}
\hspace{-1.2in}%
\includegraphics[width=.95\textwidth]{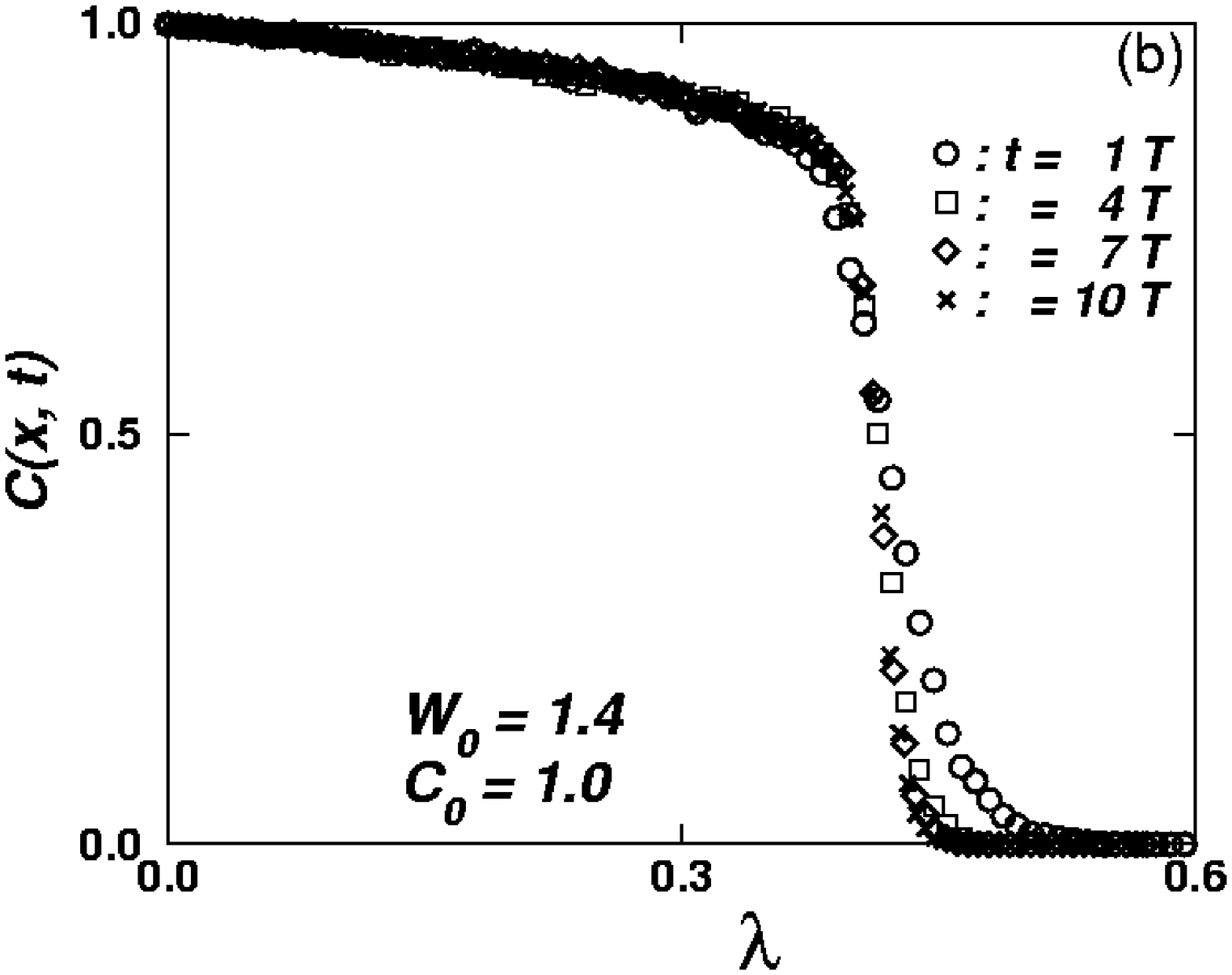}%
\raisebox{.27\textwidth}
{
\hspace{-2.7in}%
\includegraphics[width=.4\textwidth]{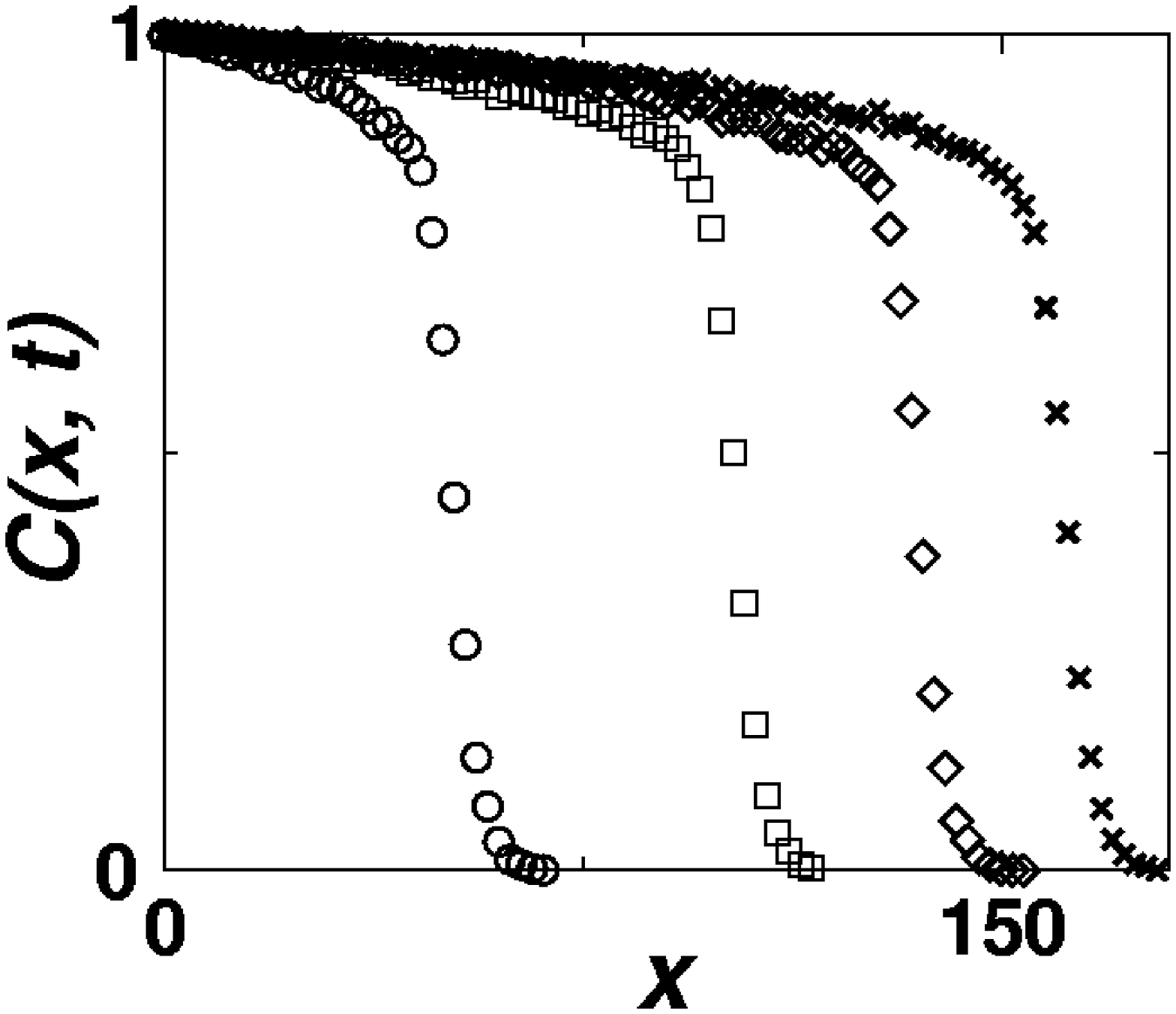}%
}%
\end{minipage}%

\vspace{.2in}

\begin{minipage}[c]{.5\textwidth}
\hspace{-.25in}%
\includegraphics[width=.95\textwidth]{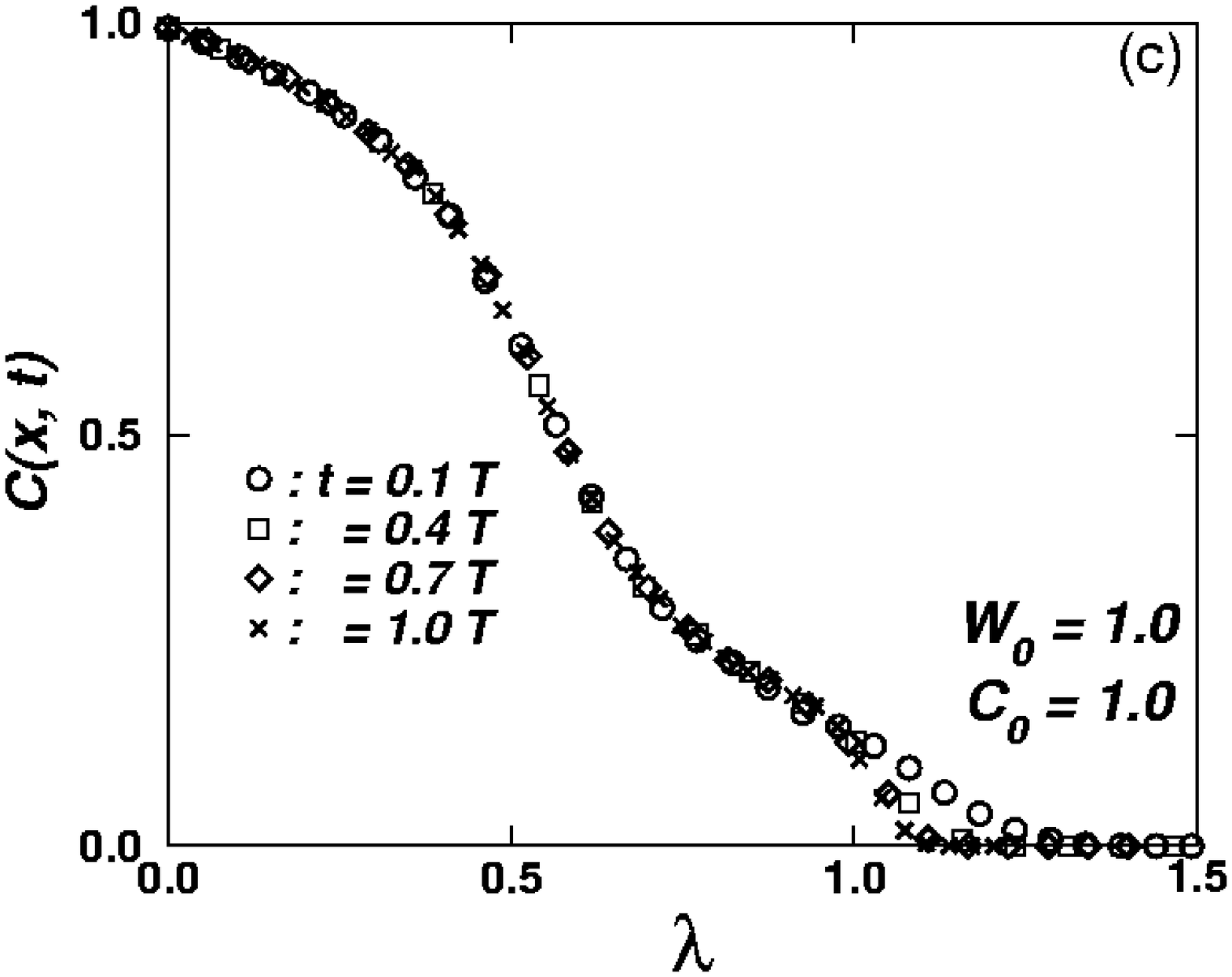}%
\raisebox{.35\textwidth}
{
\hspace{-1.7in}%
\includegraphics[width=.4\textwidth]{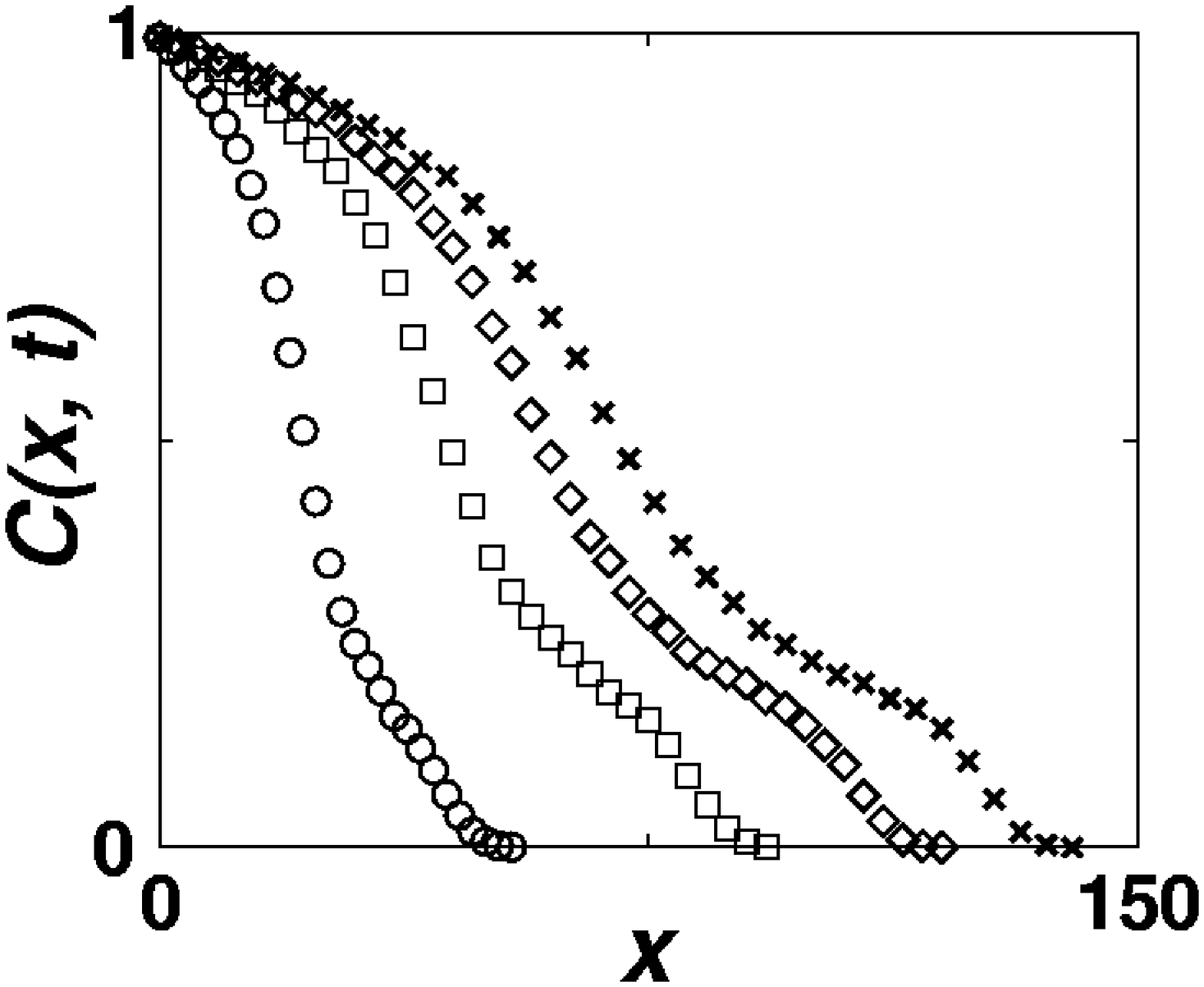}%
}%
\end{minipage}%
\begin{minipage}[c]{.5\textwidth}
\hspace{-.1in}%
\includegraphics[width=.93\textwidth]{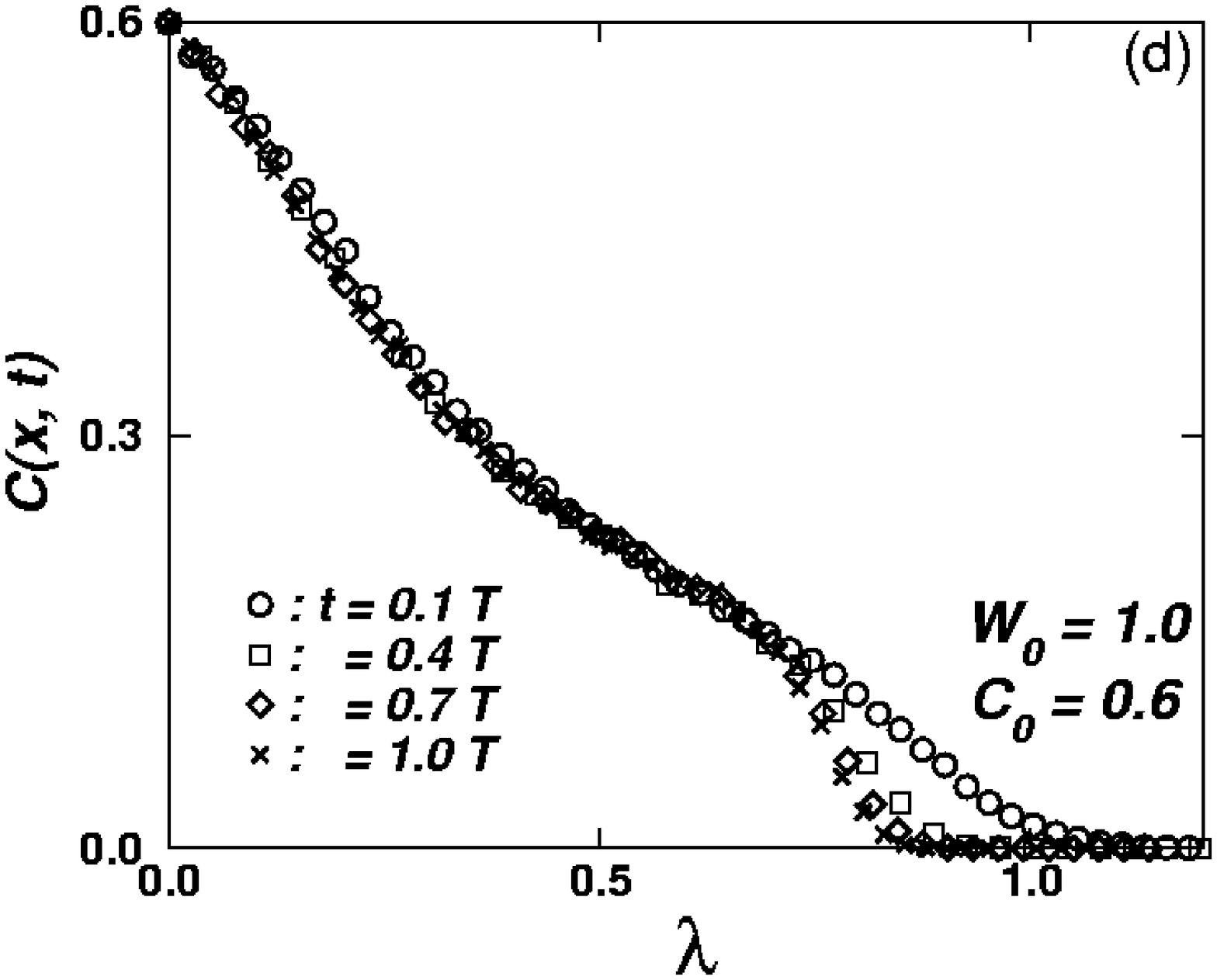}%
\raisebox{.33\textwidth}
{
\hspace{-1.6in}%
\includegraphics[width=.4\textwidth]{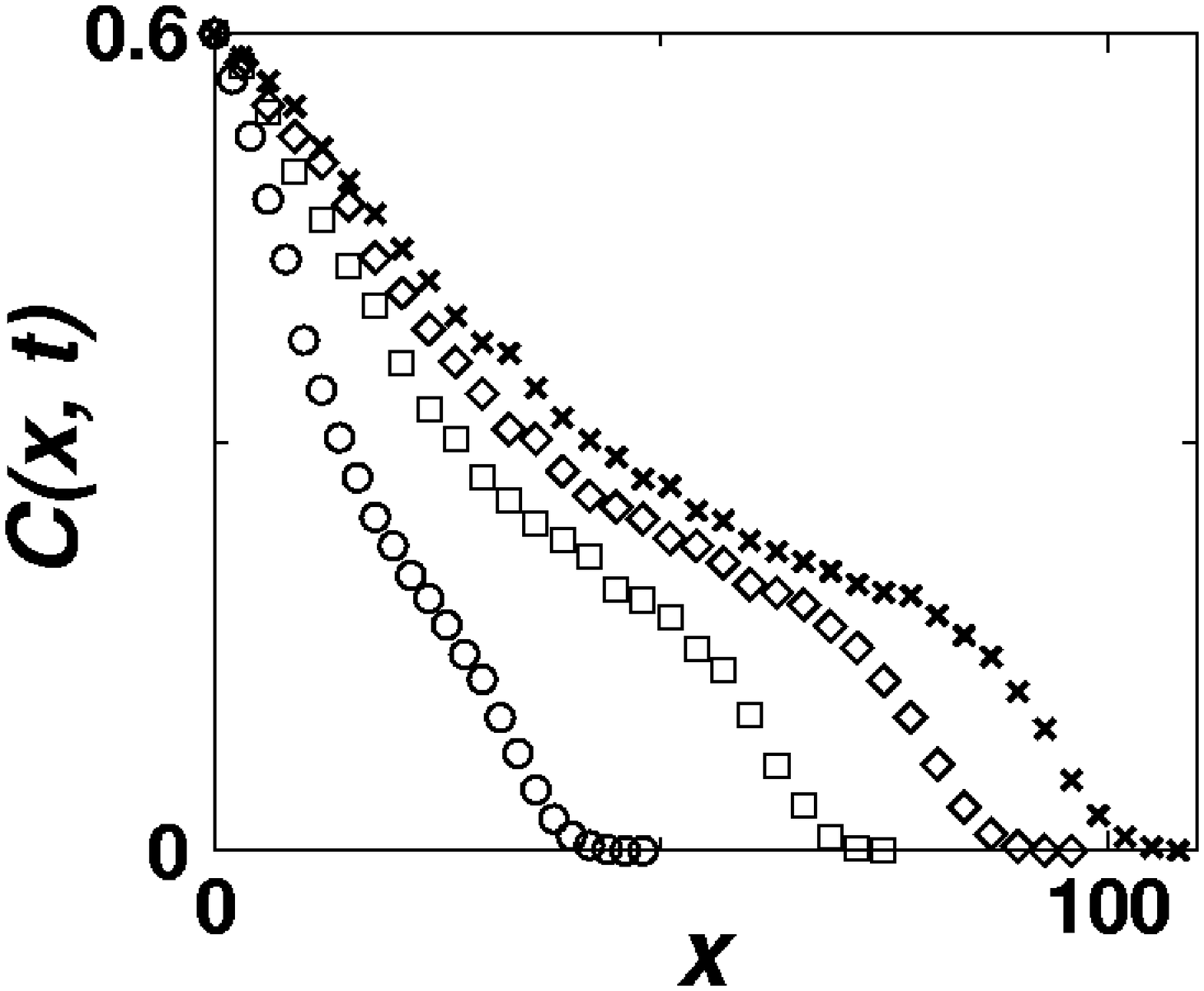}%
}%
\end{minipage}%

\begin{minipage}[c]{\textwidth}
\caption
{
\label{fig6}
Density profiles $C(x,t)$ for (a) $W_0 = 0.6,~C_0 = 1.0$;
(b) $W_0 = 1.4,~C_0 = 1.0$; (c) $W_0 = 1.0,~C_0 = 1.0$; and
(d) $W_0 = 1.0,~C_0 = 0.6$ as functions of the scaling
variable $\lambda = x/\sqrt{D_0 t}$. The insets show these
profiles as functions of $x$. The results correspond to
$t = $ $\frac{1}{10}\; T$ ($\Circle$),
$\frac{4}{10}\; T$ ($\square$), $\frac{7}{10}\; T$ ($\Diamond$),
and $T$ ($\times$) in (a), (c), and (d) and to
$t = $ $T$ ($\Circle$), $4\;T$ ($\square$), $7\;T$ ($\Diamond$),
and $10\;T$ ($\times$) in (b), where $T = 2 \times 10^6$.
The space and time units are the same as those used in
Figs.~\ref{fig2}-\ref{fig4}.
}
\end{minipage}
\end{figure}
%%%%%%%%%%%%%%
\end{widetext}

We now turn to the analysis of the time dependence of the transversally
averaged density profiles $C(x,t)$ of the spreading monolayer. Since the
time dependence of the advancing edge follows asymptotically
$X(t) \sim \sqrt{t}$ in all the cases in which spreading occurs, it is
natural to test if the density profiles $C(x,t)$ actually scale as a
function of the scaling variable $\lambda = x/\sqrt{D_0 t}$. In
Fig.~\ref{fig6} we show density profiles $C(x,t)$ for
(a) $W_0 = 0.6,~C_0 = 1.0$; (b) $W_0 = 1.4,~C_0 = 1.0$;
(c) $W_0 = 1.0,~C_0 = 1.0$; and (d) $W_0 = 1.0,~C_0 = 0.6$ as functions
of the scaling variable $\lambda = x/\sqrt{D_0 t}$, and as functions of
$x$ in the insets.
The results at late times show a very good data collapse and actually
only the data corresponding to the earliest time seems to show some
significant deviations. This strongly suggests that in the
asymptotic limit ($t \gg 1$, $X(t) \gg 1$) the density profiles can be
described by a scaling function $\tilde C(\lambda=x/\sqrt{D_0 t};W_0,C_0)$.
Here we have explicitly indicated the parametric dependence of the
scaling function on the interaction strength $W_0$ and on the density
$C_0$ at the edge of the reservoir (see Fig.~\ref{fig6}). We note that
for small $C_0$ (like $C_0 = 0.6$ in Fig.~\ref{fig6}(d)), one observes
deviations from scaling in the range of densities $C(x,t) \lesssim 0.1$.
These deviations are most probably due to insufficient statistics,
although they may also indicate that the true asymptotic regime has not
yet been reached in the simulations. However, since even in this range
there is a clear tendency of smaller changes in the shape of the density
profile for increasing time $t$, it is reasonable to expect that the
results in Fig.~\ref{fig6} are good approximations for the corresponding
scaling function $\tilde C(\lambda;W_0,C_0)$.

The scaled density profiles in Fig.~\ref{fig6} reveal three important
features. First, we have already noted that the change in shape of the
function $A(C_0)$ as $W_0$ crosses $1.0 < W_0^{(t)} < 1.2$ signals a
change in the spreading behavior. As shown by the data in
Fig.~\ref{fig6}(b), for large $W_0$ the monolayer has an almost compact
structure, and at the advancing edge there is a sharp transition from a
large density to a small, almost zero, density. Therefore, in this range
of interaction the spreading is accompanied by the emergence of a well
defined interface between two phases. In contrast, at small $W_0$ the
density decreases smoothly from the value at the edge of the reservoir
to zero and no jump in the density is visible. Thus $W_0^{(t)}$ is a
threshold value above which the attractive interaction is strong enough
to support the build-up of an interface. In this sense, the change
at $W_0^{(t)}$ may be interpreted as the onset of a phase separation.

The second point regards the shape of the density profile as a function
of the parameter $C_0$. For a weak attractive interaction $W_0$
(Fig.~\ref{fig6}(a)) or at small densities $C_0$ (Fig.~\ref{fig6}(d)),
the density profiles resemble well the error function solution of a
regular diffusion equation for non-interacting particles
\cite{Burlatsky_prl96,Oshanin_jml}. As we shall show in the next section,
in this range such a description is not only qualitatively but even
quantitatively accurate. However, for larger values $W_0$ (but still
below $W_0^{(t)}$) and for large $C_0$ (see Fig.~\ref{fig6}(c)), one
finds the formation of a pronounced shoulder in the scaling function in
the range of small $\lambda$ (i.e., $\lambda \lesssim 0.5$), and thus a
significant deviation from an error function solution. This shows that in
this range of parameters the asymmetry in the jumping probabilities due
to the attractive interaction between particles cannot be fully accounted
for by an effective boundary force approach as in
Ref.~\cite{Burlatsky_prl96,Oshanin_jml}. Therefore one has to include
explicitly this asymmetry into the description of the dynamics in order
to accurately capture the structure of the expanding film. Finally,
interaction strengths above $W_0^{(t)}$ have dramatic effects on the
spreading behavior, leading to the emergence of interfaces
(Fig.~\ref{fig6}(b)), and the simple description in terms of
non-interacting particles breaks down completely.

The third feature of the profiles to be discussed is the formation of a
``foot'' at the right end of the profile. The height of the foot is
approximately equal to $C_1$ and this is due to the fact that the density
value on an \textit{advancing} edge $\Gamma_t$ cannot decrease below $C_1$.
The formation of the step implies that the fluctuations of the interface
$\Gamma_t$ around the mean value $X(t)$ are constant or increase in time
slower than $\sqrt{t}$, such that the width of the interface divided by
$\sqrt{D_0 t}$ vanishes in the long time limit. This sharp interface is
occurring naturally due to the fact that the eventually large fluctuations
are suppressed by blocking the advancing of isolated particles ahead of the
film (see rule \textbf{(e)} in Sect. II), and thus the width of the
interface would be expected to be of the order of the cut-off $r_c = 3$
of the attractive potential and to be almost constant in time. Although the
formation of the ``foot'' is a somewhat artificial feature introduced in
the model by rule \textbf{(e)}, the advantage of it is, as discussed in
the Introduction, that it leads to the clear formation of an interface
with its associated dynamics.

\section{Continuum limit}
\subsection{Differential equation for the density and scaling behavior}
Neglecting all spatial and temporal correlations, i.e., assuming that
averages of products of occupation numbers $\eta(\bm{r};t)$ are equal
to the corresponding products of averaged occupation numbers
$\rho(\bm{r};t) = \langle \eta(\bm{r};t)\rangle$, where
$\langle \dots \rangle$ denotes the average with respect to the corresponding
probability distribution $\mathcal{P}(\{\eta(\bm{r};t)\})$ of a configuration
$\{\eta(\bm{r};t)\}$, one can formulate the following mean-field master
equation for the local occupational probability (density) $\rho(\bm{r};t)$:
%%%%%%%%%%%%%
\begin{eqnarray}
&{\displaystyle
\frac{\Delta \rho(\bm{r};t)}{\Delta t} =
-\rho(\bm{r};t) \sum_{\bm{r'},|\bm{r'}-\bm{r}|=1}
\omega_{\bm{r} \to \bm{r'};t} [1- \rho(\bm{r'};t)]
}&
\nonumber
\\
&{\displaystyle
\hspace*{.4in}+[1-\rho(\bm{r};t)] \sum_{\bm{r'},|\bm{r'}-\bm{r}|=1}
\omega_{\bm{r'} \to \bm{r};t} \rho(\bm{r'};t)\,\,,
}&
\label{master_rho}
\end{eqnarray}
%%%%%%%%%%%%
where
%%%%%%%%%%%%
\begin{equation}
U(\bm{r};t) \equiv \langle \tilde U(\bm{r};t)\rangle =
-U_0 \sum_{\bm{r''},\,0 < |\bm{r''}-\bm{r}| \leq 3}
\frac{\rho(\bm{r''};t)}{|\bm{r''} -\bm{r}|^6}
\label{potential_rho}
\end{equation}
%%%%%%%%%%%%
is replacing $\tilde U(\bm{r};t)$ in the definition for
$p(\bm{r} \to \bm{r'})$ (Eq.~(\ref{prob})).

As shown in detail in Appendix B, in the continuum space and time
limit ($\Delta t \to 0$, $a \to 0$, $\Omega^{-1} \to 0$,
$D_0 = \Omega a^2/4$ finite) of Eq.~(\ref{master_rho}), by taking
Taylor expansions for $p(\bm{r} \to \bm{r'})$ and $\rho(\bm{r'};t)$
around $\bm{r}$ and keeping terms up to second-order spatial derivatives
of the density $\rho(\bm{r};t)$ \cite{Oshanin,Lacasta,Leung}, one obtains
the following nonlinear and \textit{nonlocal} equation for $\rho(\bm{r};t)$,
%%%%%%%%%%%%
\begin{equation}
\partial_t \rho = D_0 \nabla \left[\nabla \rho +
\beta \rho \,(1-\rho) \nabla U \right]+\mathcal{O}(a^2)
\label{pde_rho}
\end{equation}
%%%%%%%%%%%%

Since the derivation of Eq.~(\ref{pde_rho}) presented in Appendix B
is not a rigorous proof (as we shall discuss below, such a proof appears
to be extremely difficult to obtain) but rather a heuristic derivation
in the spirit of Ref.~\cite{Leung}, several comments are in order before
proceeding. The only lattice gas system with long-ranged interactions
for which it has been rigorously shown that Eq.~(\ref{pde_rho}) represents
the correct continuum limit at all temperatures is the hard-core lattice
gas model with a Hamiltonian composed of short-ranged (on-site)
repulsion and long- (infinite) ranged Kac potentials evolving via rates
which satisfy detailed balance \cite{Giacomin1,Giacomin2,Giacomin3}.
Recently, it has been argued that similar equations will also hold for
systems with relatively short-ranged interactions \cite{Hildebrand,Vlachos,Lam}.
The system discussed in Refs.~\cite{Hildebrand,Vlachos,Lam} is a hard-core
lattice gas model with attractive inter-particle interaction in the form of
either a finite range constant potential or of Morse potentials, a microscopic
dynamics defined by nearest neighbor jumping rates depending on the energy at
the departure site (Arrhenius dynamics) or on the difference in energy between
the two sites (Metropolis dynamics), and a fixed density gradient imposed by fixed
densities at the boundaries of the simulation box. As shown in Ref.~\cite{Lam},
the solutions of such continuum equations compare fairly well with results of
corresponding kinetic Monte Carlo simulations in cases where the
interaction range is greater than several lattice units. Typical values
are 5-10 lattice units, depending on the dimensionality of the problem,
the type of potential, and the type of dynamics defined by the
microscopic rates. Moreover, it has been shown by Leung \cite{Leung}
that heuristic derivations, based on formal Taylor expansions, of
continuum equations from microscopic dynamics give very good results
both for the dynamics of Ising models with nearest neighbor interactions
and Kawasaki rates and for that of driven lattice gases. Based on these
results we assume that Eq.~(\ref{pde_rho}) is at least a good approximation
for the the continuum limit of our system, although it was not rigorously
derived and although the range of the attractive potential in our problem
is rather short ($r_c = 3$). Finally, since the continuum limit
for our system seems to be described by the same equation as that for the
dynamics in systems evolving via rates preserving the detailed balance
condition, one may conclude that indeed the deviations (calculated in
Appendix B) from detailed balance in the rates defined by Eq.~(\ref{rate})
are small and do not carry over to the macroscopic scale.

The constraint of a fixed density $C_0$ at the edge $x=0$ of the reservoir
implies the boundary condition
%%%%%%%%%%%%
\begin{equation}
\rho(x=0,y;t) = C_0\,.
\label{BC_x0}
\end{equation}
%%%%%%%%%%%%
As we have pointed out in Subsec.~IV.A, the condition \textbf{(e)} in the model
leads to a well defined interface and implies that for a spreading film the
minimum density on the advancing edge is $C_1 \gtrsim 0.08$. In the absence
of other additional constraints imposed by the formation of interfaces, i.e.,
for interactions $W_0 < W_0^{(t)}$, and for large times, these considerations and
the KMC results (see also Figs.~\ref{fig6}(a), (c), and (d)) strongly suggest
that the density on the advancing edge $X(t)$ can be considered as fixed and
equal to $C_1$, leading to the boundary condition
%%%%%%%%%%%%
\begin{equation}
\rho(x=X(t),y;t) = C_1\,.
\label{BC_Xt}
\end{equation}
%%%%%%%%%%%%
In what follows, we shall use the value $C_1 = 0.11$ obtained in the KMC
simulations. We note in passing that the boundary condition
Eq.~(\ref{BC_Xt}) also naturally occurred in the ``effective boundary force''
model of Burlatsky \textit{et al.} \cite{Burlatsky_prl96,Oshanin_jml}, the
expression of $C_1$ in this case being $C_1=1-\mu$, with $\mu$ the ratio
between the forward and the backward jumping rate for particles on the
advancing edge.

Since there are no boundaries along the $y-$direction and the boundary
condition at the reservoir (Eq.~(\ref{BC_x0})) is independent of $y$,
an important consequence of the $y-$ independence of Eq.~(\ref{BC_Xt})
is that the solution $\rho(\bm{r};t)$ does not actually depend on $y$,
which implies that the monolayer is homogeneous along the $y-$direction,
in agreement with the KMC results. Therefore one has to solve an
effectively one-dimensional problem. The study of the occurrence of
spontaneous transversal instabilities of the advancing edge would require
to replace Eq.~(\ref{BC_Xt}) by a moving, transversally varying boundary
condition. This might become relevant if the monolayer is driven by external
forces or encounters an obstacle. However, in view of the KMC results we
have no reason to consider such effects for the present system.

Although the reduced dimensionality is a significant simplification,
Eq.~(\ref{pde_rho}) remains quite complex because it is nonlocal due to
the term involving the interaction potential $U(\bm{r};t)$. However,
assuming that the density $\rho(\bm{r};t)$ is a slowly varying function
of the spatial coordinates (which certainly is a reasonable hypothesis
everywhere except near interfaces, see Figs.~\ref{fig2} and~\ref{fig6}),
the potential $U(\bm{r};t)$ may be expanded as
%%%%%%%%%%%%
\begin{eqnarray}
&{\displaystyle
U(\bm{r};t) = -U_0 \sum_{\bm{r'},0 < |\bm{r'}-\bm{r}| \leq 3}
\frac{\rho(\bm{r'};t)}{|\bm{r'} -\bm{r}|^6}
}&\nonumber\\
&{\displaystyle
\simeq
-U_0 \rho(\bm{r};t)\sum_{\bm{r'},0 < |\bm{r'}-\bm{r}| \leq 3}
\frac{1}{|\bm{r'} -\bm{r}|^6} + \mathcal{O}(a^2)\,.
}&
\label{expand_pot}
\end{eqnarray}
%%%%%%%%%%%%
As discussed in Appendix B, the rotational symmetry of the lattice
and of the factor $|\bm{r'}-\bm{r}|^{-6}$ implies that the summation over
$\bm{r'}$ will cancel the contributions of the first-order derivatives,
and thus the leading gradient term does not appear in the expansion above.
Because in the derivation of Eq.~(\ref{pde_rho}) only terms up to
second-order spatial derivatives of the density have been kept, i.e.,
second-order in the lattice constant $a$, and a factor $a^2$ has been already
absorbed into the diffusion coefficient $D_0$, only the zeroth-order term in the
expansion above will contribute. This leads to the \textit{local} equation
%%%%%%%%%%%%
\begin{equation}
\partial_t \rho = D_0 \nabla
\{\left[1 - g\,W_0 \rho (1-\rho)\right] \nabla \rho \} + {\cal O} (a^2),
\label{pde_rho_g}
\end{equation}
%%%%%%%%%%%%
where $g = \displaystyle{\sum_{1 \leq |\bm{r}| \leq r_c} |\bm{r}|^{-6}}$
is a geometrical factor dependent on the lattice type (e.g., square,
triangular, etc.) and on the cut-off range of the potential. For the
present case of a square lattice and a cut-off at $r_c = 3$ one has
$g \simeq 4.64$.

\begin{figure}[htb!]
\includegraphics[width=.9\columnwidth]{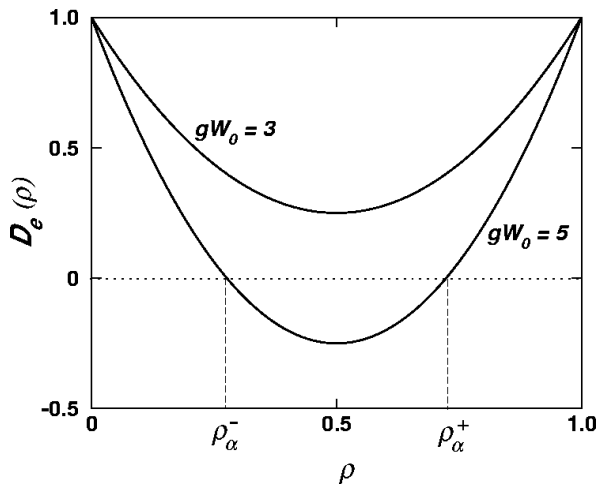}%
\caption
{
\label{fig7}
Effective diffusion coefficient $D_e(\rho)$ (Eq.~(\ref{D_rho})) for
$g W_0 = 3$ (upper curve) and $g W_0 = 5$ (lower curve), i.e., below and
above the threshold value $g W_0^{(t)} = 4$, respectively. The values
$\rho_\alpha^-$ and $\rho_\alpha^+$ indicate the range of densities
$\rho_i$ for which $D_e(\rho_i) < 0$, corresponding to instabilities in
Eq.~(\ref{diffu}).
}
\end{figure}

Rescaling the time as $t \to \tau = D_0 t$ and defining an effective
diffusion coefficient
\begin{equation}
D_{e}(\rho) = 1 - g\,W_0 \rho (1-\rho),
\label{D_rho}
\end{equation}
Eq.~(\ref{pde_rho_g}) may be written in the usual form of a diffusion
equation
%%%%%%%%%%%%
\begin{equation}
\partial_\tau \rho =
\nabla \left[D_{e}(\rho)\nabla \rho\right]+
{\cal O} (a^2) \,.
\label{diffu}
\end{equation}
%%%%%%%%%%%%
The functional form of $D_e(\rho)$ (Eq.~(\ref{D_rho})) implies that for
$W_0 > 4/g$ there will be values $\rho_i$ of the density for which
$D_{e}(\rho_i) < 0$ (see Fig.~\ref{fig7}).
%%%%%%%%%%%%%%
%%%%%%%%%%%%%%
For parameters such that $W_0 < 4/g$, Eq.~(\ref{diffu}) is a proper diffusion
equation, while for $W_0 > 4/g$ instabilities are expected in the range of
densities where $D_{e}(\rho_i) < 0$, i.e., for
$\rho_i\,\in\,\left(\rho_\alpha^-\,\,, \,\rho_\alpha^+ \right)$ where
%%%%%%%%%%%%
\begin{equation}
\rho_\alpha^\pm = \frac{1}{2}\left(1\pm\sqrt{1-\frac{4}{gW_0}}\,\,\right)\,.
\label{rho_alph_pm}
\end{equation}
%%%%%%%%%%%%%
It is known \cite{Giacomin1,Elliott} that these instabilities are
discontinuities in the density profile (``shocks''), i.e., they correspond
to the formation of interfaces, which is exactly what is observed in the KMC
results. Thus the value for the threshold interaction strength $W_0^{(t)}$
(introduced in Sec.~IV) for which interfaces emerge is predicted by the
continuum theory as $W_0^{(t)} = 4/g \simeq 0.86$. This value is significantly
smaller than the lower bound estimate $1.0 < W_0^{(t)}$ from KMC simulations,
which is not unexpected because of the mean-field character of the derivation
of the continuum equation. However, a simple, intuitive argument allows an
effective inclusion of correlations into the mean-field description and leads
to a simple correction to the mean-field value $g = 4.64$. The dynamics is
possible only by jumps into empty locations. This means that the summation
in $g$ should include at most three contributions from nearest neighbor sites,
giving $g \simeq 3.64$ and an estimate for the threshold interaction
$W_0^{(t)} \simeq 1.1$, in good agreement with the KMC results. Thus
for the rest of the analysis we will use this corrected value of $g$.

We now proceed with the analysis of the density profiles for the asymptotic
scaling limit. Since the solution $\rho(\bm{r};t)$ depends only on $x$,
Eq.~(\ref{diffu}) yields an equation for the transversally averaged density
$C(x,\tau)$\,:
%%%%%%%%%%%%
\begin{equation}
\partial_\tau C(x;\tau) =
\partial_x \left[D_{e}(C)\partial_x C(x;\tau)\right] + {\cal O} (a^2).
\label{pde_C}
\end{equation}
%%%%%%%%%%%%
Introducing the scaling variable $\lambda=x/\sqrt{\tau}$ leads to the following
equation for the scaling solution $\tilde C(\lambda)$ in the asymptotic limit
$t \gg 1$\,:
%%%%%%%%%%%%
\begin{equation}
\frac{\lambda}{2} \frac{d \tilde C}{d\lambda} +
\frac{d}{d\lambda} \left[D_{e}(\tilde C) \frac{d \tilde C}{d\lambda}\right]
+ {\cal O} [(a/\sqrt{\tau})^2] = 0
\label{ode_C}
\end{equation}
%%%%%%%%%%%%
with the boundary conditions
%%%%%%%%%%%%
\begin{subequations}
\label{BC_ode}
\begin{eqnarray}
\tilde C(0) &=& C_0\,, \label{BC_0}\\
\tilde C(A) &=& C_1\,. \label{BC_A}
\end{eqnarray}
\end{subequations}
%%%%%%%%%%%%
Since the solution of Eq.~(\ref{ode_C}) depends on whether $W_0 < W_0^{(t)}$
or $W_0 > W_0^{(t)}$, we shall discuss these two cases separately.

\subsection{Scaling solution for $W_0 < W_0^{(t)}$}
\begin{figure}[htb!]
\begin{minipage}[c]{.45\textwidth}
\includegraphics[width=.95\textwidth]{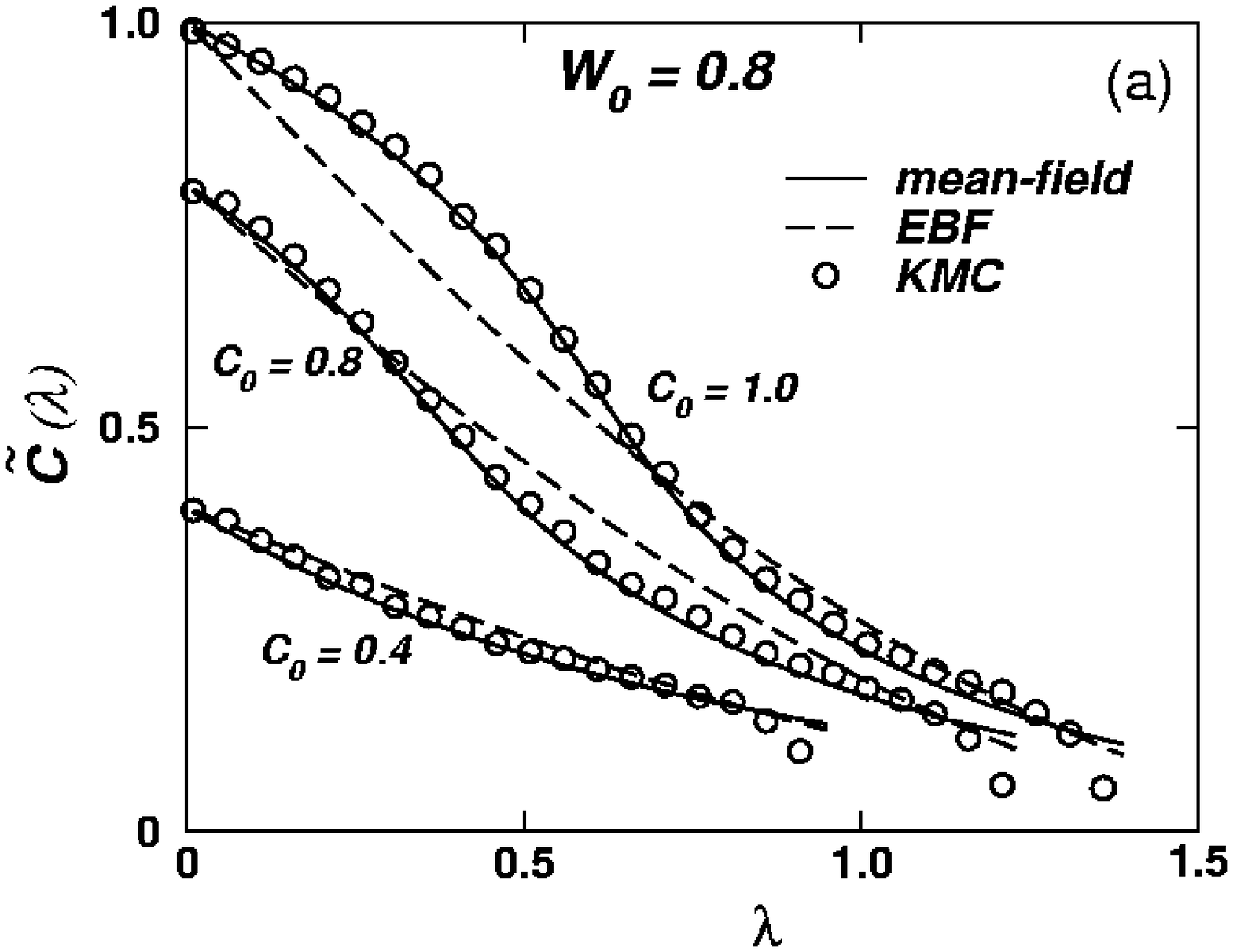}%
\end{minipage}%

\vspace{.2in}

\begin{minipage}[c]{.45\textwidth}
\includegraphics[width=.95\textwidth]{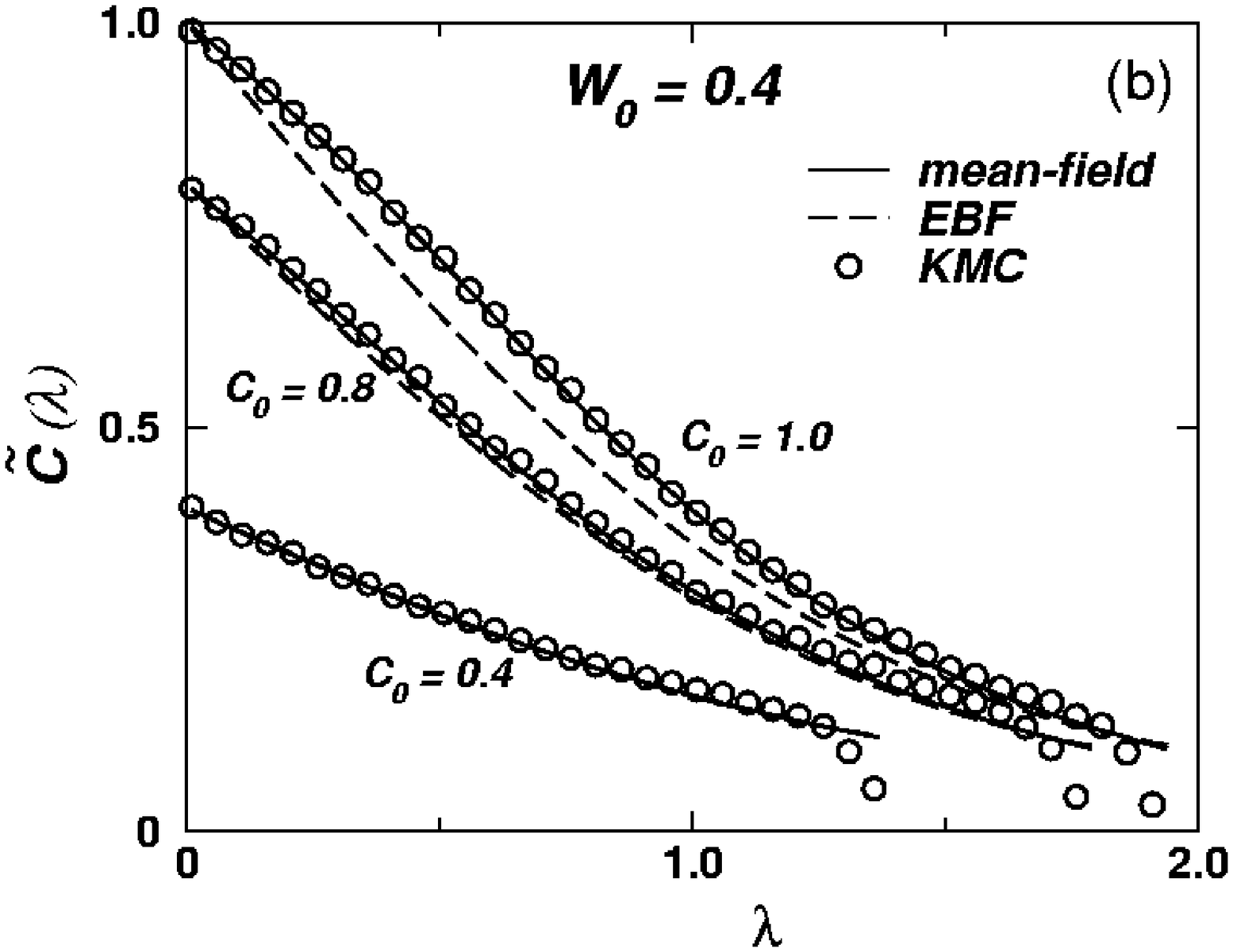}%
\end{minipage}%
\caption
{
\label{fig8}
Asymptotic scaling solution $\tilde C(\lambda)$ for
(a) $W_0 = 0.6,~\tilde C(\lambda = 0) = C_0 = 1.0,\, 0.8,\, 0.4$ and
(b) $W_0 = 1.0,~\tilde C(\lambda = 0) = C_0 = 1.0,\, 0.8,\, 0.4$ with
$\lambda=x/\sqrt{D_0 t}$. Shown are theoretical mean-field results from
Eq.~(\ref{ode_C}) (solid lines), results of the ``effective
boundary force'' (EBF) theory from Eq.~(\ref{C_OB}) (dashed lines),
and corresponding KMC results at time $T= 2 \times 10^6$ ($\Circle$)
(assumed to be close to the asymptotic limit).
}
\end{figure}

For $W_0 < W_0^{(t)}$, in Eq.~(\ref{ode_C}) the term ${\cal O} [(a/\sqrt{\tau})^2]$
may be neglected and Eq.~(\ref{ode_C}) together with the boundary conditions given
in Eq.~(\ref{BC_ode}) is a well posed problem and thus admits a regular solution
$\tilde C(\lambda;W_0,C_0)$. Although the solution cannot be found in closed form,
the numerical integration of Eq.~(\ref{ode_C}) is straightforward. Results for small
and intermediate values of the attractive coupling $W_0$ and for several values of
$C_0$ are presented in Fig.~\ref{fig8}.
%%%%%%%%%%%%%%
%%%%%%%%%%%%%%
For comparison, we also show results coresponding to the mean-field
``effective boundary force'' (EBF) approach \cite{Burlatsky_prl96} subject
to the same boundary conditions (Eq.~(\ref{BC_ode})), for which the
density profile is given by \cite{note1}
%%%%%%%%%%%%
\begin{equation}
\tilde C_{mf}(\lambda)= C_0 - (C_0 - C_1)
\frac{\textrm{erf}(\lambda/\sqrt{2})}{\textrm{erf}(A/\sqrt{2})},
\label{C_OB}
\end{equation}
%%%%%%%%%%%%
where $\textrm{erf}(z) = \frac{2}{\sqrt{\pi}}\int_0^z dy \,e^{-y^2}$
is the error function.

There is excellent agreement in all cases between the theoretical results
from Eq.~(\ref{ode_C}) and the KMC results. Similar conclusions hold for all
values of $C_0$ and $W_0 \leq 1.0$. (These results are not shown.) These findings
offer additional strong support both to the assumption that Eq.~(\ref{ode_C})
is an accurate description of the continuum limit for $W_0 < W_0^{(t)}$ and
to our heuristic correction $g \simeq 3.64$ for taking correlations into
account (see the paragraph preceding Eq.~(\ref{pde_C})). When compared to
the EBF results from Eq.~(\ref{C_OB}), we see that, as expected, at low densities
of the reservoir (see, e.g., the curves in Fig.~\ref{fig8} corresponding to
$C_0 = 0.4$) the predictions of Eq.~(\ref{ode_C}) and the EBF results
are almost identical because the monolayer is dilute and thus the
particle-particle interactions are less effective. For similar reasons, at low
values for the strength of the inter-particle attraction or at high temperature,
i.e., when $W_0$ is small, the EBF description performs well even for
high densities (see in Fig.~\ref{fig8}(b) the curve corresponding to
$C_0 = 0.8$). However, at very high densities of the reservoir or for large
values of the attractive coupling $W_0$, there are significant discrepancies
even in the \textit{qualitative} behavior between the EBF predictions
and the simulations results.
In particular, the formation of a ``shoulder'' in the case in which $W_0$ is
large (see in Fig.~\ref{fig8}(a) the curves corresponding to $C_0 = 1$ and
$C_0 = 0.8$) is remarkably well reproduced by the theoretical curve obtained from
Eq.~(\ref{ode_C}), but it is completely missed by the EBF solution (Eq.~(\ref{C_OB})).
Therefore we conclude that even in this case, i.e.,
below the threshold value $W_0^{(t)}$ for interface formation, the inter-particle
attraction has to be explicitly included into the model in order to obtain a
correct prediction for the mass distribution inside the monolayer which is extracted.

The above results should also be discussed in the context of the similar work
in Refs.~\cite{Vlachos} and \cite{Lam} mentioned in the beginning of this subsection.
We have emphasized that the derivation of the continuum limit is mean-field-like
in character, and that \textit{only} after correcting for correlations,
i.e., after adopting the improved value $g \simeq 3.64$, the continuum limit
accurately predicts both the threshold value $W_0^{(t)}$ for the interaction
coupling and the scaled asymptotic density profiles $\tilde C(\lambda;W_0,C_0)$.
Such a correction has not been included in the similar continuum equations
discussed in Refs.~\cite{Vlachos} and \cite{Lam}, and we suggest that this
explains the discrepancies observed by the authors in the case in which the
range of the inter-particle potential is short (see, e.g., the density profiles
corresponding to cut-off ranges $r_c = 2$ and $r_c = 5$ in Fig.~4(a) in
Ref.~\cite{Lam}). For the longer-ranged potentials ($r_c \geq 5$) used in
Refs.~\cite{Vlachos} and \cite{Lam}, further than nearest neighbors contribute
significantly to $g$, and thus the exclusion of a nearest-neighbor term becomes
relatively less important, which explains the good agreement obtained in the cases
$r_c \geq 5$ without any correction included.

Before proceeding to the case $W_0 > W_0^{(t)}$, we would also like to
briefly comment on the connection between our above results, the
experimental results for the precursing films of Pb on Cu(111) reported in
Ref.~\cite{Moon_01}, and the MD results for precursing films of Ag on Ni(100)
presented in Ref.~\cite{Moon_02}. The density profiles measured experimentally
(see Fig.~3(b) in Ref.~\cite{Moon_01}) and in the MD simulations (see
Fig.~3 in Ref.~\cite{Moon_02}) for the 2D spreading of Pb or Ag films show a
striking resemblance with the ones we have obtained in the KMC simulations. Moreover,
by assuming a macroscopic diffusive dynamics described by an equation of the same
form as the one derived in Eq.~(\ref{pde_C}), effective diffusion coefficients $D_e(C)$
have been obtained from the density profiles, and the data shown in Fig.~3(a) in
Ref.~\cite{Moon_01} are in qualitative agreement with a quadratic dependence for
$D_e(C)$ as in Eq.~(\ref{D_rho}). Therefore, the very simple microscopic model we
discussed seems to capture the essential features of the dynamics in these cases.
Moreover, this suggest that this form of $D_e(C)$ for Pb on Cu(111) is not necessarily
due to surface alloying -- a mechanism which is not included into the dynamics of our
model -- but is rather already a consequence of the interplay between the
concentration gradients and the inter-particle interaction, which leads to
``jamming'' and thus significantly slows down the diffusion for large values of
the density $C(x,t)$.

\subsection{Scaling solution for $W_0 > W_0^{(t)}$}
We now turn to the discussion of Eq.~(\ref{ode_C}) for the case $W_0 > W_0^{(t)}$.
Because in this case the effective diffusion coefficient $D_e(\tilde C)$ becomes
negative within a range of densities, the problem is known to be mathematically
ill-posed and to lead to discontinuities (``shocks'') in the long-time limit if
the small terms ${\cal O} [(a/\sqrt{t})^2]$ (see Eq.~(\ref{ode_C})) are set to zero
\cite{Elliott}. For this problem the existence and uniqueness of a ``weak''
solution $\tilde C(\lambda)$ (``weak'' in the sense that $\tilde C(\lambda)$
has a discontinuity at a point $\lambda = \lambda_s$ but satisfies Eq.~(\ref{ode_C})
for $\lambda \neq \lambda_s$) have been recently addressed by
Witelski \cite{Witelski1,Witelski2} using singular perturbation methods. We will use
here directly the explicit construction of the shock solution derived in
Ref.~\cite{Witelski1} for the case in which the term ${\cal O} (a^2/\sqrt{t})$
is proportional to $\partial^4_x C(x,t)$, the details of the calculation
being presented in Appendix C.

Defining
%%%%%%%%%%%%
\begin{equation}
\mu(C) = \int_0^C dC'\,D_e(C')
\label{che_pot}
\end{equation}
%%%%%%%%%%%%
Eq.~(\ref{pde_C}) may be rewritten as
%%%%%%%%%%%%
\begin{equation}
\partial_\tau C =
\partial^2_{x}\mu(C)  + {\cal O} (a^2)\,,
\label{pot_eq_C}
\end{equation}
%%%%%%%%%%%%
i.e., it has the form of a diffusion equation for $C(x,\tau)$ with a mobility
$M=1$ and a ``chemical potential'' $\mu(C)$. Moreover, as we will discuss below,
the values of $\mu(C)$ across the discontinuity satisfy conditions which are
similar to those determining the equilibrium liquid-vapor coexistence line
in the van der Waals-Maxwell mean-field theory of liquid-vapor transitions.
Because of these similarities, in what follows we shall informally denote $\mu(C)$
as chemical potential.

Following Refs.~\cite{Witelski1,Witelski2}, we look for a weak solution of
Eq.~(\ref{ode_C}), subject to the boundary conditions given in Eq.~(\ref{BC_ode}),
in the form of a ``shock'' defined as
%%%%%%%%%%%%
\begin{equation}
\tilde C(\lambda) =
\begin{cases}
C_\ell(\lambda), & ~\lambda < \lambda_s, \cr
C_r(\lambda), & ~\lambda > \lambda_s, \cr
\end{cases}
\label{shock}
\end{equation}
%%%%%%%%%%%%
where $C_\ell(\lambda)$ and $C_r(\lambda)$ satisfy Eq.~(\ref{ode_C}) in
the intervals $[0,\lambda_s)$ and $(\lambda_s,A=X(t)/\sqrt{t}]$, respectively,
subject to the boundary conditions
%%%%%%%%%%%%
\begin{eqnarray}
C_\ell(0) &=& C_0,~C_\ell(\lambda_s) = C_M,\nonumber
\\
C_r(\lambda_s) &=& C_m < C_M,~C_r(A) = C_1,
\label{BC_shock}
\end{eqnarray}
%%%%%%%%%%%%
respectively. As discussed in Appendix C, the singular perturbation analysis of
Eq.~(\ref{ode_C}) implies that the values $C_M$ and $C_m$ of the density at the
left and the right of the shock, respectively, are determined from the following
conditions expressed in terms of $\mu(C)$:
%%%%%%%%%%%%
\begin{subequations}
\label{eq_CMCm}
\begin{eqnarray}
&\mu(C_M) = \mu(C_m),& \label{contin_mu}
\\
&{\displaystyle \int_{C_m}^{C_M} dC'\,[\mu(C') - \mu(C_M)] = 0,}& \label{area_mu}
\end{eqnarray}
\end{subequations}
%%%%%%%%%%%%
i.e., continuity of the ``chemical potential'' $\mu(C)$ across the shock and a
Maxwell equal area-rule across the shock, as mentioned at the beginning of this
subsection. Solving Eq.~(\ref{eq_CMCm}), we find that the only solution satisfying
the condition $C_M > C_m$ is
%%%%%%%%%%%%
\begin{subequations}
\label{CMCm}
\begin{eqnarray}
C_M &=& \frac{1}{2} + \frac{\sqrt{3}}{2}\sqrt{1-\frac{4}{gW_0}} \,\,,\label{CM}
\\
C_m &=& \frac{1}{2} - \frac{\sqrt{3}}{2}\sqrt{1-\frac{4}{gW_0}}\,\,.\label{C_m}
\end{eqnarray}
\end{subequations}
%%%%%%%%%%%%%
Comparison with the similar expressions for $C^{\pm}_\alpha$, where
$D_e(C^{\pm}_\alpha) = 0$ (Eq.~(\ref{rho_alph_pm})), shows that for all
$W_0 > W_0^{(t)}$ one has $C_m < C^{-}_\alpha < C^{+}_\alpha < C_M$, thus
the shock occurs both above and below the interval corresponding to unstable states.
For the states corresponding to densities
$C \in {\mathfrak C} = (C_m \,, C^{-}_\alpha) \cup (C^{+}_\alpha \,, C_M)$ the
effective diffusion coefficient is positive, but the density gradients
are very large in the long-time limit and the state becomes part of the shock;
thus densities $C \in {\mathfrak C}$ correspond to metastable states.

The last unknown, the position $\lambda_s$ of the shock, is obtained from the
conservation of mass. In an infinitesimal time interval $\delta \tau$, the
displacement $\delta s = x_s(\tau+\delta\tau)-x_s(\tau)$ of the position
$x_s = \lambda_s \sqrt{\tau}$ of the shock leads to an increase
$(C_M-C_m) \delta s$ in the mass inside the stripe $x_s(\tau+\delta\tau)-x_s(\tau)$.
This should be equal to the net mass transfer
$\delta \tau [j(x_s)-j(x_s+\delta s)]$, where the mass current
$j(x) = -\partial_x \mu(C)$ (see Eq.~(\ref{pot_eq_C})) is discontinuous at $x_s$.
Since $\delta s/\delta \tau = (1/2) \lambda_s \tau^{-1/2}$ and
$\partial_x C = \tau^{-1/2} \frac{dC}{d\lambda}$, one obtains the following expression
for the position $\lambda_s$ of the shock
%%%%%%%%%%%%%%%%%%%%%%
\begin{equation}
\lambda_s = -\, 2\,\,
\frac
{
D_e(C_M){\displaystyle \left.\frac{dC}{d\lambda}\right|_{C_M}}-\,\,
D_e(C_m){\displaystyle \left.\frac{dC}{d\lambda}\right|_{C_m}}
}
{
C_M-C_m
}.
\label{lambda_shock}
\end{equation}
%%%%%%%%%%%%
We note that the above result can be also obtained via a direct integration of
Eq.~(\ref{ode_C}) across the shock, i.e., from
$\lambda = \lambda_s - \xi$ to $\lambda = \lambda_s + \xi$ in the limit $\xi \to 0$,
using for the density profile there the approximation by a step function
$C(\lambda) = C_M - (C_M-C_m) H(\lambda-\lambda_s)$, where $H(x)$ is the
Heaviside function ($H(x<0) = 0 \,, H(x\geq 0) = 1$). It is important to note here
that for sufficiently large values $W_0$ of the attractive interaction the density
$C_m$ may become smaller than $C_1$. Since the density at the advancing edge cannot be
smaller than $C_1$, in this case the branch $C_r(\lambda)$ disappears and the shock
position is obtained by setting $C_m = 0$ in Eq.~(\ref{lambda_shock}).

\begin{figure}[htb!]
\begin{minipage}[c]{.45\textwidth}
\includegraphics[width=.95\textwidth]{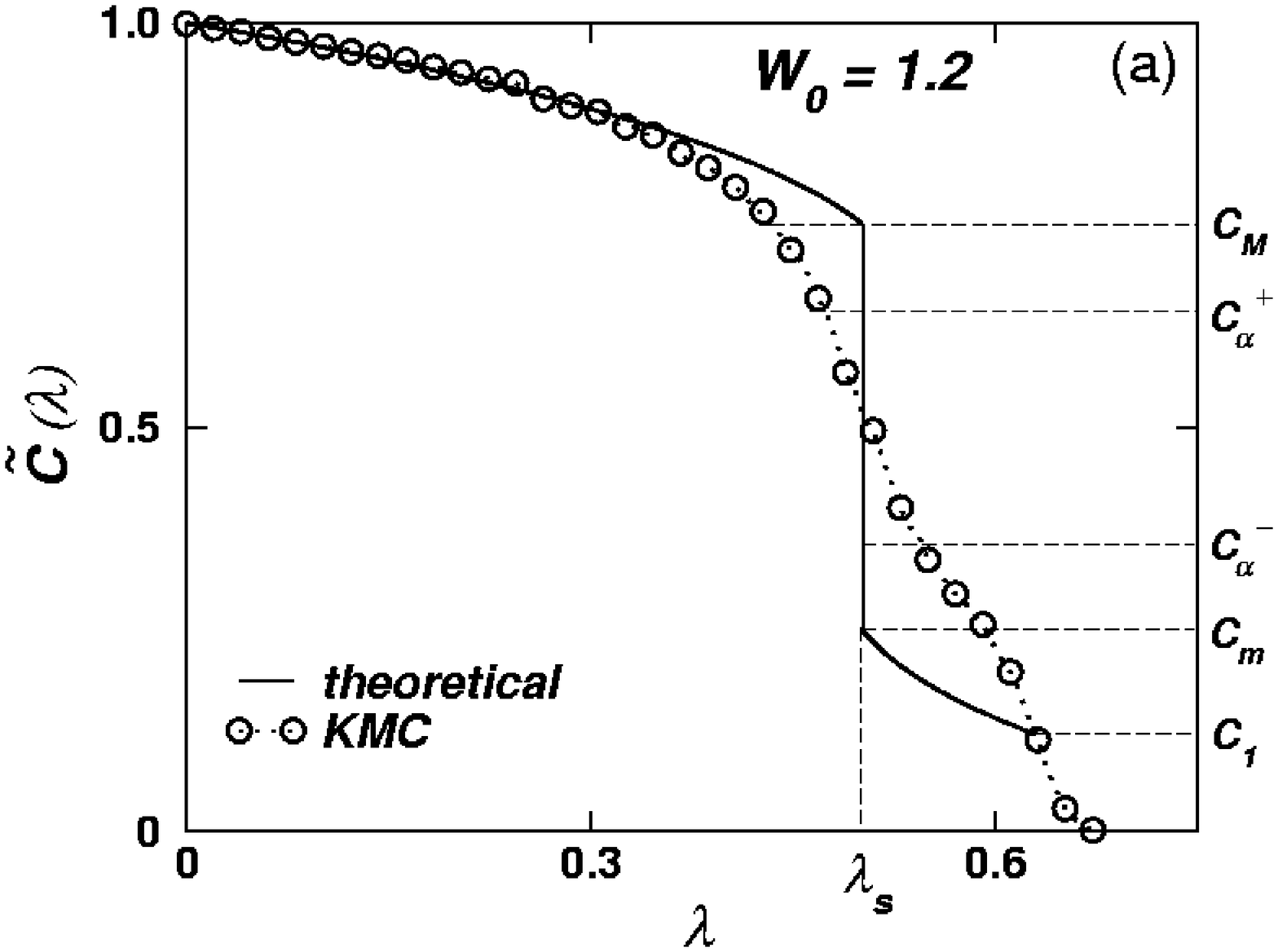}%
\end{minipage}%

\vspace{.2in}

\begin{minipage}[c]{.45\textwidth}
\includegraphics[width=.95\textwidth]{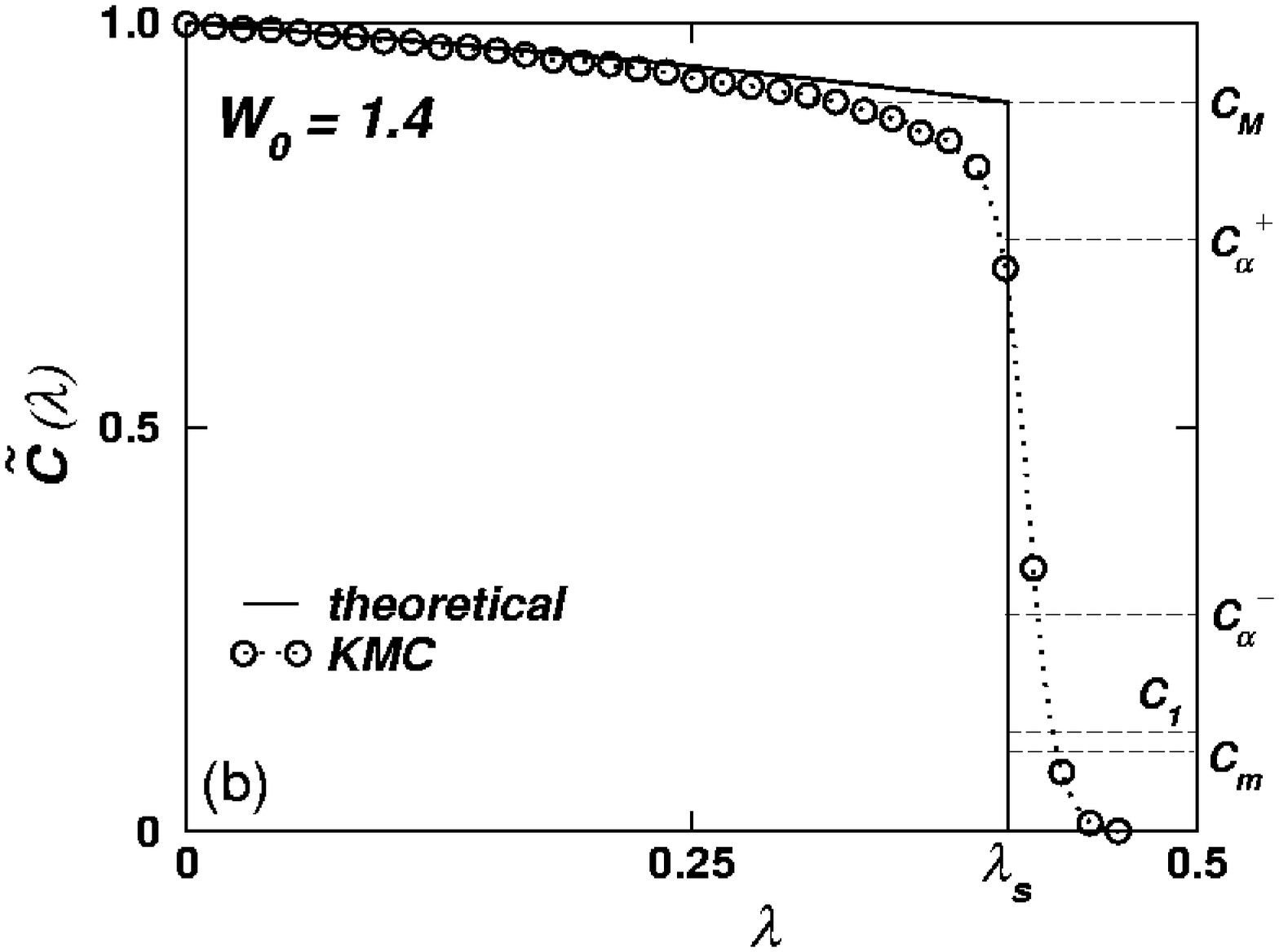}%
\end{minipage}%
\caption
{
\label{fig9}
Asymptotic scaling solution $\tilde C(\lambda)$ for (a) $W_0 = 1.2$, $C_0 = 1.0$,
and (b) $W_0 = 1.4$, $C_0 = 1.0$. Shown are theoretical results obtained from
Eqs.~(\ref{ode_C}),~(\ref{BC_shock}), and (\ref{lambda_shock}) (solid lines), and
corresponding KMC results at time $T= 2 \times 10^7$ ($\Circle$) (assumed to be
close to the asymptotic limit). The dotted line is a guide to the eye. The
dashed lines indicate the corresponding values $C_m$ and $C_M$ from Eq.~(\ref{CMCm}),
$C^\pm_\alpha$ from Eq.~(\ref{rho_alph_pm}) where $D_e(C^\pm_\alpha) = 0$, which
determines the onset of the density range leading to instabilities
(see Fig.~\ref{fig7}), $C_1$ from the boundary condition Eq.~(\ref{BC_A}),
and the position $\lambda_s$ of the discontinuity given by Eq.~(\ref{lambda_shock}).
}
\end{figure}

Once $C_m$ and $C_M$ are known, Eqs.~(\ref{ode_C}), (\ref{BC_shock}), and
(\ref{lambda_shock}) can be, in principle, solved for the corresponding quantities
$C_\ell(\lambda)$, $C_r(\lambda)$, and the position $\lambda_s$ of the shock.
Since Eq.~(\ref{ode_C}) cannot be solved in closed form, the above system of
equations has to be solved numerically. Such a numerical solution is
shown in Fig.~\ref{fig9} for the cases (a) $W_0 = 1.2$, $C_0 = 1.0$, for
which $C_m \simeq 0.25 > C_1$, and (b) $W_0 = 1.4$ and $C_0 = 1.0$, for
which $C_m \simeq 0.098 \lesssim C_1$.
%%%%%%%%%%%%%%
%%%%%%%%%%%%%%
It can be seen that the agreement between the theoretical asymptotic ``shock''
solution and the KMC measured density profiles is good for the large value
$W_0 = 1.4$, but it is not so good in the case $W_0 = 1.2$. This is very likely
due to the fact that in the latter case the simulation has not yet reached the true
asymptotic regime, while for $W_0 = 1.4$ the approach to the asymptotic shape is
faster because the low density branch $C_r$ is suppressed. In both cases
the KMC results confirm the value $C_M$ as the onset of large density gradients,
and there is good agreement between the theoretical prediction and the simulations
in the range of densities $C > C_M$. This also supports the above conclusion that
the discrepancies in the range $C < C_M$ are due to simulations times which are
not large enough.

The jump $C_M - C_m$ in the density at $\lambda_s$ explains the formation of the
plateau (for $W_0 > W_0^{(t)}$) in the dependence of $A(C_0, W_0)$ on $C_0$: if
the density $C_0$ at the reservoir is within the range $C_m \leq C_0 < C_M$, in the
immediate vicinity of the reservoir edge the density drops to $C_m$ and in the
long-time limit the extraction of the film proceeds effectively as if the reservoir
density would have been $C_m$. Also, since $1/2-C_m = -1/2 + C_M$, it follows that
the plateau should be symmetric with respect to $C_0 = 0.5$; indeed the KMC data in
Fig.~\ref{fig4}(a) exhibit this symmetry (as long as $W_0$ is such that
$C_m > 0.1$). Moreover, since the density must satisfy $C \leq 1$, one may
conclude that for interaction values $W_0$ such that $C_M > 1$ the extraction of
a monolayer is no longer possible. This implies that the exact value for the upper
limit of the interaction $\widetilde W_0^{(cov)}$ above which no macroscopic film
is extracted from the reservoir is given by $\widetilde W_0^{(cov)} = 6/g \simeq 1.65$.
This value is significantly below the value
$\widetilde W_0^{(cov)} = W_0^{(cov)}(C_0 = 1) \simeq 2.3$ extracted from the
linear extrapolation of the KMC data (see Fig.~\ref{fig4}(b)), the discrepancy
very likely reflecting that the KMC simulations have not yet reached the true
asymptotic limit (or that a linear extrapolation is not appropriate). Thus it is to
be expected that in the range $C_0 \gtrsim 0.85$ the separatrix $W_0^{(cov)}$ shown
in Fig.~\ref{fig4}(b) significantly overestimates the correct curve.

\section{Summary and Conclusions}

Using kinetic Monte Carlo (KMC) simulations and a non-linear diffusion equation within
the continuum limit, we have studied a lattice-gas model with interacting
particles for the two-dimensional spreading on homogeneous substrates of a fluid
monolayer which is extracted from a reservoir (Fig.~\ref{fig1}).
We have obtained the following main results:

1. The two-dimensional KMC simulations confirm the time dependence
$X(t \to \infty) = A \sqrt t$ of the spreading, where $X(t)$ is the average position of
the advancing edge of the monolayer at time $t$, and reveal a non-trivial dependence of
the prefactor $A$ on the strength $U_0$ of inter-particle attraction and on the fluid
density $C_0$ at the reservoir (see Figs.~\ref{fig2}, \ref{fig3}, and \ref{fig4}).
A careful analysis of this behavior has allowed us to identify, in terms of
$W_0 = U_0/k_B T$, a transition point $W_0^{(t)} \simeq 1.1$ associated with the occurrence
of interfaces inside the extracted monolayer, and to estimate a ``covering phase-diagram''
in the $W_0 \textrm{--} C_0$ plane (Fig. \ref{fig4}) together with a
``covering -- non-covering'' separatrix $W_0^{(cov)}$ below which a macroscopic
film is extracted from the reservoir, while above $W_0^{(cov)}$ it is not
extracted.

2. The asymptotic (i.e., at long time and large spatial scales) transversally averaged
density profiles $C(x,t)$ measured in the KMC simulations exhibit a scaling behavior
as function of the scaling variable $\lambda = x/\sqrt{D_0 t}$, where
$D_0$ is the one-particle diffusion coefficient on the bare substrate (Fig.~\ref{fig6}).
They clearly show that for this model the density in the extracted monolayer is not
spatially constant, in contrast to the predictions of other theoretical models mentioned
in the Introduction. This provides an unambiguous -- and otherwise difficult -- way to
experimentally discriminate between the various theoretical models proposed. Moreover,
the simulations show that the present model predicts qualitatively different structures
for the experimentally accessible density profiles above and below the threshold value
$W_0^{(t)}$ (see Figs.~\ref{fig7} and \ref{fig8}), in particular the formation of sharp
interfaces inside the extracted monolayer for $W_0 > W_0^{(t)}$.

3. The asymptotic, scaled density profiles $\tilde C (\lambda)$ have been analyzed within
a continuum limit with the corresponding non-linear diffusion equation derived from the
microscopic master equation. Within this approach we have included the effect of
correlations in an effective manner into the standard mean-field description by adapting
the value of the integrated attractive interaction to account for the presence of empty
nearest-neighbor sites (see Fig.~\ref{fig5}). This leads to an excellent agreement between
the theoretical predictions based on the continuum limit and the KMC results both for
the value $W_0^{(t)}$ and for the scaled density profiles (Fig.~\ref{fig8}).
Additionally we have shown that, even below the threshold value $W_0^{(t)}$ for interface
formation, the inter-particle attraction has to be explicitly included into the model
in order to obtain correct predictions for the mass distribution inside the extracted
monolayer. The formation of the interfaces in the range $W_0 > W_0^{(t)}$ has been
related to instabilities of the diffusion equation associated with densities for which
the corresponding effective diffusion coefficient becomes negative (Fig.~\ref{fig7}).
We have constructed the corresponding discontinuous density profiles (``shocks'') and
critically compared them with the KMC measured ones (Fig.~\ref{fig9}). Based on the results
of a singular perturbation analysis, we have obtained a good estimate
$\widetilde W_0^{(cov)} \simeq 1.65$ for the upper limit of the interaction above which no
macroscopic film is extracted from the reservoir.

Finally, we comment on the connection between this model and experimental systems.
As briefly discussed in Subsect.V.B, the present model appears to provide a successful
description for the diffusion of solid metals on metal surfaces as studied in
Refs.~\cite{Moon_01} and \cite{Moon_02}. We have found a \textit{qualitative} agreement
between the experimental results in the case of diffusion of Pb on Cu(111) \cite{Moon_01}
and our theoretically derived density profiles and effective diffusion coefficient.
It seems to be promising to explore \textit{quantitatively} the applicability of the present
model for such metal on metal systems. To this end the experimental setup described in
Refs.~\cite{Moon_01} and \cite{Moon_02} would have to be modified in order to have straight
instead of circular spreading geometries and a deposit-substrate combination chosen such as
to avoid surface alloying effects.

As noted in the Introduction the experiments with fluids performed so far deal with polymer
oils. As long as the entanglement of the polymer chains is not important, one may consider a
coarse-grained description in which the chain is replaced by an effective particle of the
size of the corresponding radius of gyration and only the motion of the center of mass is
considered. Although the motion of these effective particles might not resemble simple,
activated jumping processes so that the microscopic model description is not directly
applicable, it is reasonable to expect that the macroscopic evolution will be diffusive.
Therefore, one may expect that the continuum limit of the present model can be used to
describe the spreading behavior observed in experiments with polymer oils, but the
macroscopic parameters entering into the diffusion equation should be regarded as fit
parameters and not quantities calculated from microscopic dynamics as considered here.

In order to obtain a direct, quantitative test of the present theoretical predictions
for precursor liquid films, new experiments would have to be performed using simple liquids
chosen such that they have a spreading rate large compared with the evaporation rate. This
should be combined with observation techniques chosen such that the density profiles, and
not only the spreading rate, could be measured, which would require an in plane (lateral)
resolution in the order of few nm for the case of simple liquids, i.e., several lattice
constants, and in the order of 10-50 nm for the case of polymer oils, i.e., several
inter-``effective'' particles distances, because the density variations are expected to
occur on larger length scales. One technique which possibly may fulfill
these requirements is reflection interference contrast microscopy \cite{sackmann},
assuming that the microscope objective may scan the area of interest (of the order of
$\mathrm{mm}^2$) in times sufficiently small compared to those on which the density
profile changes. The technique has been used before in studies of (equilibrium) wetting
properties on micropatterned solid surfaces \cite{sackmann}, and already at the time of
its first implementation a lateral resolution of at least 200 nm (see Fig.~13(b) in
Ref.~\cite{sackmann}) combined with a normal resolution of the order of 1 nm has been
achieved.

%%%%%%%%%%%%
\begin{acknowledgments}
This work has been supported by the Deutsche Forschungsgemeinschaft
within the priority program ``Wetting and Structure Formation at
Interfaces'', Grant DI 315/7-3.
\end{acknowledgments}

%%%%%%%%%%%%
\appendix

\section{Kinetic Monte Carlo Method}

In this section we discuss in some detail the variable-step
continuous-time kinetic Monte Carlo algorithm
\cite{Binder,Adam_KMC,Jansen} that we have used. The main idea
is to consider the sequence of independent, uncorrelated events
represented by jumps of particles away from the wells in which
they were residing. Each of these events has an identical
time- and environment-independent rate $\Omega$ as shown by
Eq.~(\ref{total_rate}), in contrast to the location-dependent
rates of particular transitions $\bm{r} \to \bm{r'}$
(Eq.~(\ref{rate})).

Consider the system at time $t$ when there are $N$ particles in
the film ($x > 0$) and an event just occurred. Since the attempts
of any particle to leave its well are uncorrelated to similar
events of other particles, and since for each particle the rate
for a successful jump is $\Omega$, for each one of the particles
the probability that until time $t' > t$ no successful attempt
occurs is $P_1(\tau) = \exp(-\Omega \tau)$, where $\tau = t'-t$.
Therefore, since the jumps are uncorrelated, the probability
that \textit{none} of the $N$ particles experienced a successful
attempt in $\tau$ is
$P_N(\tau) = [P_1(\tau)]^N = \exp(- N \Omega \tau)$ and the
probability that the first successful jump will take place at $t'$
will be given by
\begin{equation}
P = N \Omega P_N(\tau) = N \Omega \exp(- N \Omega \tau).
\label{tau}
\end{equation}
Thus the time interval $\tau$ between successful jumping attempts
(between ``events") is a random variable distributed according to
Eq.~(\ref{tau}). Since all the $N$ particles have identical rates
$\Omega$ for events, the probability for a certain particle to be
the one undergoing the jump is $\Omega/(N\Omega) = 1/N$, i.e., the
particle to jump is selected at random. Let us assume the selected
particle is at location $\bm{r}$. There are $z = 4$ nearest-neighbor
locations, and thus $z = 4$ possible realizations of the jump;
the one to be actually attempted is selected according to the
probability defined by Eq.~(\ref{prob}). Specifically, calling the
four probabilities $p_1,\dots,p_4$, with $p_1$ corresponding to the
jump $(x,y) \to (x+1,y)$ and the others being indexed counter
clockwise, one compares the successive sums
$s_0 \equiv 0$, $\displaystyle{s_j=\sum_{k=1}^{k=j} p_k}$,
$j=1,2,3,4$, with a random number $v \in [0,1]$ and selects $p_k$ for
which $s_{k-1} < v \leq s_k$. As described in the text, the jump
takes place if the selected destination site is empty, and is rejected
if the destination site is occupied.

We note here that, as shown in Ref.~\cite{Adam_KMC}, incrementing
the time between events using intervals generated according to
Eq.~(\ref{tau}) and not a constant time interval equal to the average
time $1/(N\Omega)$ between events, like in a classical Monte Carlo
simulation, is essential in assuring that the simulated time is the
correct real time, and thus that the simulations capture the correct
time development of spreading.

\section{Heuristic Derivation of The Continuum Limit}

The (mean-field) master equation for the local occupational probability
(density) $\rho(\bm{r};t)$ is given by
%%%%%%%%%%%%%
\begin{equation}
\begin{split}
\frac{\partial \rho(\bm{r};t)}{\partial t} =
-\rho(\bm{r};t) \sum_{\bm{r'},|\bm{r'}-\bm{r}|=1}
\omega_{\bm{r} \to \bm{r'};t} [1- \rho(\bm{r'};t)]\\
+[1-\rho(\bm{r};t)] \sum_{\bm{r'},|\bm{r'}-\bm{r}|=1}
\omega_{\bm{r'} \to \bm{r};t} \rho(\bm{r'};t),
\end{split}
\label{master}
\end{equation}
%%%%%%%%%%%%
where
%%%%%%%%%%%%
\begin{equation}
\label{ome}
\omega_{\bm{r} \to \bm{r'};t} =
\Omega
\frac
{
\displaystyle{\exp \biggl\{ \frac{\beta}{2}
\left[U(\bm{r};t)-U(\bm{r'};t)\right]\biggr\}}
}
{
\displaystyle{\sum_{\bm{r'}, |\bm{r'} -\bm{r}|=1}}
\exp \biggl\{ \frac{\beta}{2}
\left[U(\bm{r};t)-U(\bm{r'};t)\right]\biggr\}
}
\end{equation}
%%%%%%%%%%%%
and
%%%%%%%%%%%%
\begin{equation}
U(\bm{r};t) =
-U_0 \sum_{\bm{r''},\,0 < |\bm{r'}-\bm{r}| \leq 3}
\frac{\rho(\bm{r''};t)}{|\bm{r} -\bm{r''}|^6}.
\label{pot}
\end{equation}
%%%%%%%%%%%%%%%%%%

We consider a two-dimensional regular lattice of coordination number $z$
and lattice constant $a$ and choose the orthogonal $x-y$ coordinate system
such that the $x-$axis is along one of the lattice directions. For a
given site $\bm{r}$ we index the nearest neighbors as $\bm{r'}_j$,
$j = 0,1,2,...z-1$, where $j = 0$ is chosen such that $\bm{r'}_0 - \bm{r}$
is parallel to the $x$-axis and $j$ runs in counterclockwise direction.
Denoting the angle formed by the vector $\bm{r'}_j - \bm{r}$ with the
$x$-axis as $\phi_j$ so that $\phi_j = \displaystyle{2 \pi\frac{j}{z}}$,
the components $x'_j$ and $y'_j$ of $\bm{r'}_j - \bm{r}$ are given by
$x'_j = a \cos(\phi_j)$ and $y'_j = a \sin(\phi_j)$. The following
relations are satisfied by the angles $\phi_j$ \cite{gradshteyn} and will
prove to be useful for the rest of the calculation:
%%%%%%%%%%%%
\begin{subequations}
\label{sums}
\begin{eqnarray}
&
\displaystyle{
\sum_{j=0}^{z-1}\sin^k(\phi_j) = 0,~
\sum_{j=0}^{z-1}\cos^k(\phi_j) = 0,
}& \nonumber\\
&
k~\textrm{odd,}~ 0< k < z,
&\label{sum_k}
\\
&
\displaystyle{
\sum_{j=0}^{z-1}\sin^2(\phi_j) = \frac{z}{2},~
\sum_{j=0}^{z-1}\cos^2(\phi_j) = \frac{z}{2},
}
&\label{sum_2}
\\
&
\displaystyle{
\sum_{j=0}^{z-1}\sin(2\phi_j) = 0,~
\sum_{j=0}^{z-1}\cos(2\phi_j) = 0,
}
&\label{sum_twice}
\end{eqnarray}
\end{subequations}
%%%%%%%%%%%%%%%%%%

Defining
\begin{equation}
\delta h(\bm{r},\bm{r'}_j;t) = h(\bm{r'}_j;t)-h(\bm{r};t),
\label{delta}
\end{equation}
where $h(\bm{r};t)$ is any of the functions $\rho(\bm{r};t)$, $U(\bm{r};t)$,
or products of them, expanding $h(\bm{r'}_j;t)$ near $\bm{r}$, and
summing $\delta h(\bm{r},\bm{r'}_j;t)$ over $\bm{r'}_j$ one obtains
%%%%%%%%%%%%%%%%%%
\begin{eqnarray}
&&\sum_{j=0}^{z-1} \delta h(\bm{r},\bm{r'}_j;t) =
a \left[
\frac{\partial h}{\partial x} \sum_{j=0}^{z-1}\cos(\phi_j)+
\frac{\partial h}{\partial y} \sum_{j=0}^{z-1}\sin(\phi_j)
\right]
\nonumber\\
&&+\frac{a^2}{2}
\left[
\frac{\partial^2 h}{\partial x^2}
\sum_{j=0}^{z-1}\cos^2(\phi_j)+
2 \frac{\partial^2 h}{\partial x \partial y}
\sum_{j=0}^{z-1}\sin(\phi_j) \cos(\phi_j)\right.
\nonumber\\
&& \left.+ \frac{\partial^2 h}{\partial y^2}
\sum_{j=0}^{z-1}\sin^2(\phi_j)
\right]+\dots~.
\end{eqnarray}
%%%%%%%%%%%%%%%%%%%
In the relation above, $h \equiv h(\bm{r};t)$ and the derivatives are
evaluated at $\bm{r}$. Replacing the corresponding sums by the results
in Eq.~(\ref{sums}), it follows that
%%%%%%%%%%%%%%%%%%%
\begin{equation}
\sum_{j=0}^{z-1} \delta h(\bm{r},\bm{r'}_j;t) =
\frac{z a^2}{4} \nabla^2 h + {\cal O}(a^4).
\label{sum_delta}
\end{equation}
%%%%%%%%%%%%%%%%%%%
Straightforward algebra allows one to derive from the definition
(\ref{delta}) the following additional useful relations involving a second
function $f(\bm{r};t)$:
%%%%%%%%%%%%
\begin{subequations}
\label{sum_delta_prod}
\begin{eqnarray}
&f(\bm{r'}_j;t) \delta h(\bm{r},\bm{r'}_j;t) =
\delta (fh)(\bm{r},\bm{r'}_j;t)
&
\nonumber\\
&
- h(\bm{r}_j;t) \delta f(\bm{r},\bm{r'}_j;t)
&
\label{sum_func_delta}
\\
&
\delta f(\bm{r}, \bm{r'}_j;t) \delta h(\bm{r},\bm{r'}_j;t) =
\delta (fh)(\bm{r},\bm{r'}_j;t)
&
\nonumber\\
&-h(\bm{r}_j;t) \delta f(\bm{r},\bm{r'}_j;t)-
f(\bm{r}_j;t) \delta h(\bm{r},\bm{r'}_j;t).
&
\label{sum_delta_delta}
\end{eqnarray}
\end{subequations}
%%%%%%%%%%%%%%%%%%

Assuming that $U(\bm{r};t)$ varies slowly on the scale of the lattice constant
so that $\beta \delta U(\bm{r},\bm{r'};t) \ll 1$, one has the expansion
%%%%%%%%%%%%%%%%%%
\begin{equation}
\begin{split}
\frac
{
{
\displaystyle
\exp \left[ -\frac{\beta}{2} \,
\delta U(\bm{r},\bm{r'}_k;t)\right]
}
}
{
{
\displaystyle
\sum_{j=0}^{z-1}
\exp \left[ -\frac{\beta}{2} \,
\delta  U(\bm{r},\bm{r'}_j;t)\right]
}
}\simeq\hspace{.2\columnwidth}\\
\simeq
\frac
{
{
\displaystyle
\exp \left[ -\frac{\beta}{2} \,
\delta U(\bm{r},\bm{r'}_k;t)\right]
}
}
{
{
\displaystyle
\sum_{j=0}^{z-1}
\left\{1-\beta \,\delta U(\bm{r},\bm{r'}_j;t)/2+
\left[\beta \,\delta U(\bm{r},\bm{r'}_j;t)/2\right]^2 +\dots
\right\}
}
}.
\end{split}
\label{prob_expand}
\end{equation}
%%%%%%%%%%%%%%%%%%
Thus using Eqs.~(\ref{sum_delta}) and (\ref{sum_delta_delta}) one obtains
%%%%%%%%%%%%
\begin{equation}
\begin{split}
\frac{\omega_{\bm{r} \to \bm{r'};t}}{\Omega}
\simeq
\frac
{1}{z} \,
{\displaystyle
\exp \left[ -\frac{\beta}{2} \,
\delta U(\bm{r},\bm{r'}_k;t)\right]
}\times\hspace{.1in}
\\
\times{\displaystyle
\left\{1+
\frac{\beta a^2}{8} \,
\left[\nabla^2 U-\beta (\nabla U)^2\right]+{\cal O}(a^4)
\right\}
},
\end{split}
\label{prob_a_powers}
\end{equation}
where as before $U \equiv U(\bm{r};t)$ and the spatial derivatives are
evaluated at $\bm{r}$. As we shall show below, the zeroth-order term in
the above expansion (Eq.~(\ref{prob_a_powers})) contributes to the master
equation already in the order $a^2$, and therefore the other terms on the
RHS of Eq.~(\ref{prob_a_powers}) will lead to contributions proportional to
$a^4$ and higher orders. Therefore, up to contributions which are of second
order in the lattice constant in the master equation,
Eq.~(\ref{prob_a_powers}) may be rewritten as
%%%%%%%%%%%%%%%%%%%
\begin{equation}
\omega_{\bm{r} \to \bm{r'};t} \simeq
\frac{\Omega}{z}
{\displaystyle
\exp \left[ -\frac{\beta}{2} \,
\delta U(\bm{r},\bm{r'};t)\right]
}.
\label{p_asquared}
\end{equation}
%%%%%%%%%%%%%%%%%%
This means that the deviations of the rates $\omega_{\bm{r} \to \bm{r'};t}$
from detailed balance, which according to Eq.~(\ref{prob_a_powers}) are
of second order and higher in the lattice constant, in the equation corresponding
to the continuum limit contribute with terms of fourth order and higher in
the lattice constant. These terms become negligible in the limit $a \to 0$.

The expression (\ref{p_asquared}) may be now expanded in terms of powers of
$\beta \delta U(\bm{r},\bm{r'};t)$:
%%%%%%%%%%%%%%%%%%%
\begin{equation}
\omega_{\bm{r} \to \bm{r'};t} \simeq p_0 + p_1 + p_2 + \dots \,,
\label{p_deltaU}
\end{equation}
%%%%%%%%%%%%%%%%%%
where
%%%%%%%%%%%%
\begin{subequations}
\label{prob_terms}
\begin{eqnarray}
p_0 &=& \frac{\Omega}{z},
\label{p0}
\\
p_1 &=& -\frac{\Omega \beta}{2z} \,\delta U(\bm{r},\bm{r'};t),
\label{p1}
\\
p_2 &=& \frac{\Omega \beta^2}{4z}\, [\delta U(\bm{r},\bm{r'};t)]^2.
\label{p2}
\end{eqnarray}
\end{subequations}
%%%%%%%%%%%%%%%%%%
Note that the expression (\ref{p_asquared}) implies
%%%%%%%%%%%%%%%%%%%
\begin{eqnarray}
\omega_{\bm{r'} \to \bm{r};t} &\simeq&
\frac{\Omega}{z}
{\displaystyle
\exp \left[ -\frac{\beta}{2} \,
\delta U(\bm{r'},\bm{r};t)\right]
}
\nonumber\\
&=&
\frac{\Omega}{z}
{\displaystyle
\exp \left[ +\frac{\beta}{2} \,
\delta U(\bm{r},\bm{r'};t)\right]
}
\label{p_reversed}
\end{eqnarray}
%%%%%%%%%%%%%%%%%%
and thus its expansion in powers of $\beta \delta U(\bm{r},\bm{r'};t)$ is
%%%%%%%%%%%%%%%%%%%
\begin{equation}
\omega_{\bm{r'} \to \bm{r};t} \simeq p_0 - p_1 + p_2 + \dots \,.
\label{p_rev_deltaU}
\end{equation}
%%%%%%%%%%%%%%%%%%

We now will compute separately the contributions of these terms to the RHS
of Eq.~(\ref{master}). For the contribution due to $p_0$ one has
%%%%%%%%%%%%%
\begin{eqnarray}
\textrm{RHS}_{(0)} &=& \frac{\Omega}{z} \,
{
\displaystyle
\sum_{j=0}^{z-1}
\left\{
\rho(\bm{r'}_j,t)[1-\rho(\bm{r},t)]
\right.
}\nonumber\\
&-&
\left.
\rho(\bm{r},t)[1-\rho(\bm{r'}_j,t)]
\right\}
\nonumber\\
&=&
\frac{\Omega}{z} \,
{
\displaystyle
\sum_{j=0}^{z-1}\delta \rho(\bm{r},\bm{r'}_j;t)
}
\nonumber\\
&=&
\frac{\Omega a^2}{4} \nabla^2 \rho + \Omega \,{\cal O}(a^4)\,.
\label{rhs0}
\end{eqnarray}
%%%%%%%%%%%%
For $p_1$ one has
%%%%%%%%%%%%%
\begin{eqnarray}
\textrm{RHS}_{(1)} &=& \frac{\beta \Omega}{2z} \,
{
\displaystyle
\sum_{j=0}^{z-1} \delta U(\bm{r},\bm{r'}_j;t)
\left\{
\rho(\bm{r'}_j,t)[1-\rho(\bm{r},t)]
\right.
}
\nonumber\\
&+&
\left. \rho(\bm{r},t)[1-\rho(\bm{r'}_j,t)]
\right\}
\nonumber\\
&=&
\frac{\beta \Omega}{2z}\, [1-2\rho(\bm{r};t)]
{
\displaystyle
\sum_{j=0}^{z-1}\rho(\bm{r'}_j;t)\delta U(\bm{r},\bm{r'}_j;t)
}
\nonumber\\
&+&
\frac{\beta \Omega}{2z}\, \rho(\bm{r};t)
{
\displaystyle
\sum_{j=0}^{z-1}\delta U(\bm{r},\bm{r'}_j;t).
}
\label{rhs1_part}
\end{eqnarray}
%%%%%%%%%%%%
Using Eqs.~(\ref{sum_delta}) and (\ref{sum_func_delta}) to replace the two
sums in the expression above, one obtains
\begin{eqnarray}
&&\textrm{RHS}_{(1)} = \frac{\beta \Omega a^2}{8} \,
\left\{
(1-2\rho)[\nabla^2 (\rho U) - U \nabla^2 \rho] +
\rho \nabla^2 U
\right\} \nonumber\\
&& +\,\Omega {\cal O}(a^4)
= \frac{\beta \Omega a^2}{4} \,
\nabla[\rho (1-\rho)\nabla U] + \Omega {\cal O}(a^4)\,.
\label{rhs1}
\end{eqnarray}
%%%%%%%%%%%%
Finally, for $p_2$ one has
%%%%%%%%%%%%%
\begin{eqnarray}
\textrm{RHS}_{(2)} &=& \frac{\beta^2 \Omega}{4z} \,
{
\displaystyle
\sum_{j=0}^{z-1} [\delta U(\bm{r},\bm{r'}_j;t)]^2
\left\{
\rho(\bm{r'}_j,t)[1-\rho(\bm{r},t)]
\right.
}
\nonumber\\
&-&
\left. \rho(\bm{r},t)[1-\rho(\bm{r'}_j,t)]
\right\}
\nonumber\\
&=&
\frac{\beta^2 \Omega}{4z} \,
{
\displaystyle
\sum_{j=0}^{z-1}[\delta U(\bm{r},\bm{r'}_j;t)]^2 \delta \rho(\bm{r},\bm{r'}_j;t)
}\,.
\label{rhs2_part}
\end{eqnarray}
%%%%%%%%%%%%
Using repeatedly Eqs.~(\ref{sum_func_delta}) and (\ref{sum_delta}) in the
above sum, one obtains
\begin{eqnarray}
\textrm{RHS}_{(2)} &=& \frac{\beta^2 \Omega a^2}{4} \,
[\nabla^2 (\rho U^2) - \rho \nabla^2 U^2 - U^2 \nabla^2 \rho
\nonumber\\
&-& 2 U \nabla^2 (\rho U) + 2 U \rho \nabla^2 U + 2 U^2 \nabla^2 \rho]
\nonumber\\
&+& \Omega {\cal O}(a^4) = \Omega {\cal O}(a^4)\,.
\label{rhs2}
\end{eqnarray}
%%%%%%%%%%%%
It is easy to see that higher order terms, $p_3, \dots$, will contribute
with terms which are at least of the order $a^4$, and thus the expressions
(\ref{rhs0}) and (\ref{rhs1}) are the only terms relevant for Eq.~(\ref{master}).

Collecting the terms and passing to the continuum limit
$a \to 0$, $\Omega^{-1} \to 0$ such that $D_0 = \Omega a^2/4$ stays finite,
one arrives at the result given in Eq.~(\ref{pde_rho}) in the main text, i.e.,
%%%%%%%%%%%%
\begin{equation}
\partial_t \rho = D_0 \nabla \{ \nabla \rho +
\beta \left[\rho (1-\rho) \nabla U \right] \}+ \mathcal{O}(a^2)\,.
\label{cont_limit}
\end{equation}
%%%%%%%%%%%%

\section{Derivation of The Shock Solution}

Following Ref.~\cite{Witelski1}, we start from equation Eq.~(\ref{pot_eq_C})
written in terms of the scaling variable $\lambda$ as
%%%%%%%%%%%%
\begin{equation}
-\frac{1}{2}\lambda \frac{dC}{d\lambda} = \frac{d^2}{d\lambda^2} \mu(C)  +
\left(\frac{a}{\sqrt{\tau}}\right)^2
Q\left[\frac{d^4C}{d\lambda^4},\left(\frac{d^2C}{d\lambda^2}\right)^2,\dots\right]
\,,
\label{equ_C_e2}
\end{equation}
%%%%%%%%%%%%
where the function $Q$ is a linear combination of fourth-order derivatives terms
of the form $\frac{d^4C}{d\lambda^4}$, $\left(\frac{d^2C}{d\lambda^2}\right)^2$,
$\left[\frac{d^2 (UC)}{d\lambda^2}\right]^2,\dots$.
The region of interest, $\lambda \in [0,A]$, naturally decomposes into the region
near the interface, $\lambda_s - h(\epsilon) \leq \lambda \leq \lambda_s +
h(\epsilon)$, and the outer region $|\lambda-\lambda_s| > h(\epsilon)$ with
$\epsilon = a/\sqrt{\tau}$; $h(\epsilon)$ is a smooth function such that
$h(\epsilon \to 0) = 0$ (which ensures that in the long-time limit the width of
the interface becomes negligible) and
$\lim_{\epsilon \to 0} \frac{h(\epsilon)}{\epsilon} \to \infty$, i.e., it is
assumed that the decrease of the width is slower than $\epsilon$. In the outer
region, the solution $C_{l,r}(\lambda)$ is a slowly varying, smooth function of
$\lambda$, and the terms proportional to $\epsilon^2$ in Eq.~(\ref{equ_C_e2}) are
negligible. In contrast, in the inner region the gradients are very large, and the
fourth-order terms become relevant.

In order to obtain the shock structure, we change to the ``stretching''
variable $\zeta = (\lambda - \lambda_s)/\epsilon$ for
$|\lambda-\lambda_s| \leq h(\epsilon)$ and look for a smooth, strictly decreasing
solution $C(\zeta)$. In terms of $\zeta$ Eq.~(\ref{pot_eq_C}) turns into
%%%%%%%%%%%%
\begin{equation}
\begin{split}
-\frac{1}{2}(\lambda_s + \zeta \epsilon)\epsilon^{-1}\frac{dC}{d\zeta} = \hspace{.6in}\\
\hspace{.1in}
\epsilon^{-2}\frac{d^2}{d\zeta^2} \mu(C)  +  \epsilon^{-2}
Q\left[\frac{d^4C}{d\zeta^4},\left(\frac{d^2C}{d\zeta^2}\right)^2,\dots\right]
\,,
\end{split}
\label{shock_zeta_first}
\end{equation}
%%%%%%%%%%%%
which in the limit $\epsilon \to 0$ leads to the zeroth order in the
$\epsilon$-approximation:
%%%%%%%%%%%%
\begin{equation}
\frac{d^2}{d\zeta^2} \mu(C)  +
Q\left[\frac{d^4C}{d\zeta^4},\left(\frac{d^2C}{d\zeta^2}\right)^2,\dots\right]
\, = 0.
\label{shock_zeta}
\end{equation}
%%%%%%%%%%%%
Since in the limit $\epsilon \to 0$ the inner region
$-h(\epsilon)/\epsilon \leq \zeta \leq h(\epsilon)/\epsilon$ is mapped into
$-\infty < \zeta < +\infty$ and at the ends of the shock region the inner solution
must match the outer solution $C_{l,r}(\lambda)$, Eq.~(\ref{shock_zeta}) should be
solved subject to the boundary conditions
%%%%%%%%%%%%
\begin{equation}
\lim_{\zeta \to -\infty} C(\zeta) = C_M \,\,,\,\,
\lim_{\zeta \to +\infty} C(\zeta) = C_m,
\label{BC_shock_zeta}
\end{equation}
%%%%%%%%%%%%
which implies also that all the derivatives $C^{(k)}(\zeta)$ of the
smooth inner solution $C(\zeta)$ tend to zero as $\zeta \to \pm \infty$.

Although the function $Q$ may be computed explicitly by using Eq.~(\ref{sum_delta}),
%%%%%%%%%%%%%%%%%%%
\begin{equation}
\sum_{j=0}^{z-1} \delta h(\bm{r},\bm{r'}_j;t) =
\frac{z a^2}{4} \nabla^2 h + \frac{3z a^4}{8} \nabla^4 h + {\cal O}(a^6),
\label{sum_delta_a4}
\end{equation}
%%%%%%%%%%%%%%%%%%%
and following similar steps as in calculating the terms proportional to $a^2$ in
Appendix B, the result is very complicated and a singular perturbation analysis
appears to be extremely difficult, if at all possible. However, one may argue that
the term $\partial^4_x C$ is always relevant in the region of the shock because it
is associated with an interface contribution ${\cal F}_i = \int dx [\nabla C(x,t)]^2$
to the free energy functional ${\cal F} = {\cal F}_i + \cdots$ of a dynamical
Cahn-Hillard theory of phase separation,
$\partial_t C =
\nabla\left[M(C)\nabla\left(\frac{\delta {\cal F}}{\delta C}\right)\right]$ (where $M(C)$
denotes the mobility)\cite{Witelski2}. Therefore, all the other terms that are relevant
should be of the same order as $\partial^4_x C$. This leads us to the approximation
%%%%%%%%%%%%
\begin{equation}
Q\left[\frac{d^4C}{d\zeta^4},\left(\frac{d^2C}{d\zeta^2}\right)^2,\dots\right]
\simeq q \frac{d^4C}{d\zeta^4},
\label{Q_app}
\end{equation}
%%%%%%%%%%%%
with $q$ a constant or a very slowly varying function of $\lambda$, e.g.,
$q(\zeta) = \tilde q(\lambda_s + \epsilon \zeta)$. In this case,
Eq.~(\ref{shock_zeta}) reduces to
%%%%%%%%%%%%
\begin{equation}
\frac{d^2}{d\zeta^2} \left[\mu(C) + q \frac{d^2C}{d\zeta^2}\right] = 0.
\label{shock_conserv}
\end{equation}
%%%%%%%%%%%%
A first integration leads to
$\frac{d}{d\zeta} \left[\mu(C) + q \frac{d^2C}{d\zeta^2}\right] = const = 0$
because both $\frac{d}{d\zeta} \mu(C)$ and $\frac{d^3C}{d\zeta^3}$ are zero at
$\zeta \to \pm \infty$, and thus a second integration yields
%%%%%%%%%%%%
\begin{equation}
\mu(C) + q \frac{d^2C}{d\zeta^2} = b,
\label{shock_final}
\end{equation}
%%%%%%%%%%%%
where $b$ is an integration constant. Since
$\lim_{\zeta\to\pm\infty} \frac{d^2C}{d\zeta^2} = 0$, one obtains
$\mu[C(\zeta\to\pm\infty)] = b$, i.e., the requirement of continuity of
$\mu(C)$ across the shock (Eq.~(\ref{contin_mu}) in the main text):
%%%%%%%%%%%%
\begin{equation}
\mu(C_M) = \mu(C_m)\,\,.
\label{conti}
\end{equation}
%%%%%%%%%%%%
Since $\frac{dC}{d\zeta} \neq 0$ (except at infinity) and $b = \mu(C_M)$,
Eq.~(\ref{shock_final}) may be rewritten in the form
%%%%%%%%%%%%
\begin{equation}
\frac{q}{2} \frac{d}{d\zeta}\left[\left(\frac{dC}{d\zeta}\right)^2\right] =
[\mu(C_M)-\mu(C)] \frac{dC}{d\zeta} \,\,,
\label{shock_area}
\end{equation}
%%%%%%%%%%%%
which leads to
%%%%%%%%%%%%
\begin{equation}
\int_{C_m}^{C_M} dC [\mu(C_M)-\mu(C)] =
\frac{q}{2} \int_{+\infty}^{-\infty} d\zeta
\frac{d}{d\zeta}\left[\left(\frac{dC}{d\zeta}\right)^2\right] = 0\,\,,
\label{area_rule}
\end{equation}
%%%%%%%%%%%%
i.e., the equal area-rule for $\mu(C)$ (Eq.~(\ref{area_mu}) in the main text).

Finally, we remark that all the details of the calculation, as well as the main
results (Eqs.~(\ref{conti}) and (\ref{area_rule})), remain unchanged if the
corrections $Q$ would have the form $Q[C^{(4)},(C^{(2)})^2,\dots] =
\frac{d^2}{d\zeta^2}P[C, C^{(2)},(C^{(1)})^2,\dots]$, with P a linear combination of
terms of second order spatial derivatives satisfying
$\lim_{\zeta \to \pm \infty} P \to 0$ \cite{note2}, and thus it seems reasonable to assume
that in general it is \textit {only} the inner structure of the shock that depends
on the particular form of $Q$ \cite{Witelski1,Witelski2}.

\end{document}